\documentclass[a4paper,11pt]{article}
\pdfoutput=1 

\usepackage{jheppub} 

\usepackage[T1]{fontenc} 

\usepackage{float}
\usepackage{amsmath,amssymb,graphics,epsfig,subfigure}
\usepackage{color}
\usepackage{hyperref}
\usepackage{epstopdf}
\usepackage{graphicx}

\newcommand \beq{\begin{equation}}
\newcommand \eeq{\end{equation}}
\newcommand \beqn{\begin{eqnarray}}
\newcommand \eeqn{\end{eqnarray}}
\newcommand \bseq{\begin{subequation}}
\newcommand \eseq {\end{subequations}}
\newcommand \nn{\nonumber}

\newcommand \hmn{h_{\mu\nu}}
\newcommand \mn{{\mu\nu}}

\title{\boldmath Smooth braneworld in $6$-dimensional asymptotically AdS spacetime}

\author[a,b]{Jun-Jie Wan,}
\author[a,b]{Zheng-Quan Cui,}
\author[a,b]{Wen-Bin Feng}
\author[a,b,c,1]{and Yu-Xiao Liu \note{Corresponding author.}}

\affiliation[a]{Institute of Theoretical Physics and Research Center of Gravitation, Lanzhou University,\\Lanzhou 730000, P.R. China}
\affiliation[b]{Lanzhou Center for Theoretical Physics and Key Laboratory of Theoretical Physics of Gansu Province, Lanzhou University, Lanzhou 730000, P.R. China}
\affiliation[c]{Key Laboratory for Magnetism and Magnetic Materials of the MOE, Lanzhou University,\\Lanzhou 730000, P.R. China}

\emailAdd{liuyx@lzu.edu.cn}

\abstract{In this paper, we investigate a six-dimensional smooth thick braneworld model which contains a compact extra dimension and an infinite large one. The braneworld is generated by a real scalar field with a $\phi^6$ potential and the bulk is an asymptotically $\text{AdS}_6$ spacetime. The geometry achieves the localization of the free $U(1)$ gauge field, which is a problem in five-dimensional Randall-Sundrum-like models. In addition, we analyze the stability of the braneworld system and the localization of the graviton.}

\begin{document}
\maketitle
\flushbottom

\section{Introduction}

As an extra-dimensional theory, the braneworld scenario provides an alternative {framework} for outstanding issues, such as the hierarchy problem  \cite{ArkaniHamed:1998rs,Antoniadis:1998ig,Randall:1999ee,Randall:1999vf} and the cosmological constant problem \cite{Weinberg:1988cp,Kim:2000mc}. Hence, the braneworld theory has attracted extensive attention and various braneworld models have been proposed.

The braneworld theory must be consistent with experiments. The Newtonian potential indicates there are only three large spatial dimensions. Therefore, it is natural to compact extra dimensions into a small spatial volume, which needs high energy to observe. One of the famous theories is the Kaluza-Klein (KK) theory \cite{Kaluza:1921tu,Klein:1926tv}. However, Rubakov and Shaposhnikov proposed the possibility of noncompact extra dimensions, known as the domain wall model \cite{Rubakov:1983bb,Akama:1982jy}. The model shows that we live inside a domain wall generated by a scalar field in a five-dimensional flat spacetime, but the Newtonian potential is not recovered. Later, Randall and Sundrum (RS) proposed the RS-2 model where the Newtonian potential can be recovered on the brane even though the extra dimension is infinite \cite{Randall:1999vf}. Various extensions of the RS-2 model {have been} investigated  \cite{Csaki:2000fc,Gremm:1999pj,Gherghetta:2000qt,Arias:2002ew,Afonso:2006gi,Liu:2011am,Agashe:2014jca}.

The RS-2 model assumes that particles in the standard model (SM) of particle physics are localized on a hypersurface (3-brane) {in} the bulk \cite{Randall:1999vf}. A more realistic braneworld model should provide a reasonable dynamic interpretation for this assumption. Combining the RS-2 model and the domain wall {model}, {braneworld models} with thickness and inner structure were proposed as thick {branes} \cite{Csaki:2000fc,DeWolfe:1999cp,Gremm:1999pj,Shiromizu:1999wj}. In this scenario, various matter fields distribute in the bulk. {The {SM} fields corresponding to {zero modes} of various bulk matter fields and interactions should be reproduced on the brane~\cite{Smolyakov:2015zsa}, 
so it is important to investigate the localization of various bulk matter fields.}

{For} five-dimensional RS-like braneworld models, ref. \cite{Bajc:1999mh} indicates that not all free boson fields are localized on the brane. {It shows} that a free massless scalar field and {graviton} are localized but a free $U(1) $ gauge field is not. {Hence}, the localization mechanisms of gauge fields have been widely studied \cite{Kehagias:2000au,Oda:2001ux,Dvali:2000rx,Batell:2006dp,Flachi:2009uq,Chumbes:2011zt,Cruz:2012kd,Vaquera-Araujo:2014tia,Alencar:2014moa,Zhao:2014gka,Zhao:2014iqa,Zhao:2017epp,Zhou:2017bbj,Freitas:2018iil}. The usual way to localize the $U(1)$ gauge field is {adding couplings} \cite{Kehagias:2000au,Oda:2001ux,Dvali:2000rx,Batell:2006dp,Chumbes:2011zt,Vaquera-Araujo:2014tia,Cruz:2012kd,Zhao:2014gka,Zhou:2017bbj}.
For these mechanisms, some questions have been raised in ref. \cite{Freitas:2018iil}. For example, some {additional couplings} are included in refs. \cite{Oda:2001ux,Dvali:2000rx,Batell:2006dp,Vaquera-Araujo:2014tia}, but it is not clear what the meaning of the {additional couplings is.} Do such mechanisms work for other braneworld models? {Therefore}, refs. \cite{Zhao:2017epp,Flachi:2009uq,Alencar:2014moa,Zhao:2014iqa,Freitas:2018iil,Sui:2020fty} proposed geometrical coupling mechanism{s} by introducing the coupling terms $R A_M A^M$ and $R_{MN} A^M A^N$, and found that {the unique condition} for localization is that the bulk is asymptotically AdS \cite{Freitas:2018iil}. For geometrical coupling mechanisms, we still have some questions. What do these coupling terms mean? These coupling terms have effect on the four-dimensional effective action. Whether the {four-dimensional effective} theory {coincides} with observations, especially in a strong gravity region? In view of all this, we will consider the minimal coupling between the $U(1)$ gauge field and gravity in this paper.

In ref. \cite{Freitas:2018iil}, the authors found that the localization condition for the free $U(1)$ gauge field in an {asymptotically} AdS spacetime is $(D-4) > d$, where $D$ is {the number of the bulk dimensions} and $d$ is the number of {infinite} extra dimensions. Note that, for a fixed $d$, the condition $(D-4) > d$ can always be satisfied by introducing compact extra dimensions. In this case{,} the matter fields need to be localized {not} on the 3-brane but on a $(D-d)$-dimensional submanifold containing these compact dimensions. For example, six-dimensional models with an infinite extra dimension and a compact one satisfy {this} condition.

The types of six-dimensional braneworld models can be classified by the geometry of the extra-dimensional space (the transverse space) {which may}
\begin{itemize}
\item {be} two infinitely large dimensions
    \cite{Silva:2017crq,Giovannini:2001hh,Kanno:2004nr,Vinet:2004bk},
\item be a 2-sphere \cite{Olasagasti:2000gx}, a football {shape }\cite{Garriga:2004tq}, or an apple {shape} \cite{Gogberashvili:2007gg},
\item be a torus \cite{Duan:2006es,Liu:2007gk},
\item include a conical deficit   \cite{Chodos:1999zt,VazquezPoritz:2001zt,Brummer:2005sh,Firouzjahi:2005qs,Noguchi:2005ws,Navarro:2003vw,Papantonopoulos:2007fk,Kogan:2001yr},
    {and}
\item be a cylinder \cite{Oda:2000zc,Oda:2000zj,Gherghetta:2000jf,Sousa:2014dpa}.
\end{itemize}
The models with a cylinder-like transverse space whose geometry satisfies the localization condition of free vector fields are studied in refs. \cite{Gherghetta:2000jf,Oda:2000zj,Oda:2000zc,Sousa:2014dpa}. {They suggest} that the matter fields are localized on a $4$-brane {denoted by $5$-diemnsional manifold $\mathcal{M}_4 \times \mathcal{S}_1$, while the $3$-brane corresponds to $\mathcal{M}_4$}.

The consistency conditions of localization of the six-dimensional free $U(1)$ gauge field have been investigated in refs. \cite{Fu:2018erz,Freitas:2020mxr,Freitas:2020vcf}, which must be satisfied for a six-dimensional braneworld model.
The geometry in ref. \cite{Oda:2000zc} satisfies the localization condition of the free vector field but is not generated by dynamical fields. The aim of this paper is to propose a braneworld model which satisfies both the consistency conditions \cite{Fu:2018erz} and the localization conditions \cite{Freitas:2018iil} of the free $U(1)$ gauge field. We expect that our model has the following properties.
\begin{itemize}
\item It is a thick brane model which includes a compact extra dimension and an infinite one.
\item The brane is generated by a dynamical field.
\item The spacetime is stable.
\item The Newtonian potential can be recovered.
\item To be consistent with observations, the compact extra dimension should be sufficiently small.
\end{itemize}

This paper is organized as follows.
In section \ref{Setup}, we introduce the setup of our model. In section \ref{Brane Solution in 6D Spacetime}, we present a new brane solution.
In section \ref{linear perturbations and localization}, the tensor mode of gravitational perturbations is studied.
In section \ref{Localization of Fields}, the localization of boson fields is explored.
The last section \ref{Conclusions and Outlook} is devoted to conclusions and outlook.



\section{Setup}
\label{Setup}

We consider a six-dimensional braneworld model constructed by a real scalar field $\phi$, and start with the action
\begin{eqnarray}\label{action}
	S =   \frac{{M_{*}^4}}{2} \int d^6 x
	\sqrt{-g} \left(R
	+ \mathcal{L}_{\text{m}}\right),
	\label{1}
\end{eqnarray}
where $g=\mathrm{det}(g_{MN})$ with $M$ and $N$ denoting six-dimensional spacetime indices{. Here} {$M_{*}$} is the six-dimensional fundamental energy scale. We use the unit $M_*=c=\hbar=1$ in {the following}.
The Lagrangian of the background scalar field is
\begin{equation}\label{Lagrangian of the background scalar}
\mathcal{L}_{\text{m}} =  - \frac{1}{2}g^{MN}{\partial_M}\phi {\partial _N}\phi  - {V_{\Lambda}}(\phi ),
\end{equation}
where
\begin{eqnarray}
	V_{\Lambda}(\phi)&=&V(\phi)+\Lambda.
\end{eqnarray}
Here we write the six-dimensional cosmological constant $\Lambda$ independently. From the Lagrangian \eqref{Lagrangian of the background scalar}, {the energy-momentum tensor is}
\begin{eqnarray}
	T_{MN}{\equiv}-\frac{2}{\sqrt{-g}}\frac{\delta\left(\sqrt{-g}\mathcal{L}_{\text{m}}\right)}{\delta g^{MN}}
	=\partial_M\phi\partial_N\phi+g_{MN}\mathcal{L}_\text{m}.
\end{eqnarray}
The variation of the action \eqref{action} with respect to the metric $g_{MN}$ yields the field equation
\begin{equation}\label{BS-EinsteinEquations}
	{R_{MN}} - \frac{1}{2}{g_{MN}}R ={T_{MN}},
\end{equation}
and the equation of motion for the scalar field $\phi$ is
\begin{equation}
\square^{(6)} \phi=\frac{\partial {V_{\Lambda}}}{\partial \phi},\label{eqphi}
\end{equation}
where
$\square^{(6)}=g^{M N} \nabla_{M} \nabla_{N}$ is the six-dimensional d'Alembert operator.

In this work, we are interested in a six-dimensional spacetime $\mathcal{M}_4\times {\mathcal{R}_1} \times \mathcal{S}_1$, where $\mathcal{M}_4$ is a four-dimensional Minkowski manifold and ${\mathcal{R}_1} \times \mathcal{S}_1$ is a transverse manifold.
The metric ansatz can be written as
\begin{eqnarray}
	ds^2=a^2(y)\eta_{\mu\nu} dx^{\mu}dx^{\nu}+dy^2+b^2(y)R_0^2d\theta^2,
	\label{metric}
\end{eqnarray}
where $y \in (-\infty,\infty$) denotes the infinite extra dimension and $\theta \in [0,2\pi)$ denotes the compact extra dimension with radius $R_0$. Similar to {the} KK theory, we assume that $R_0$ is {sufficiently} small, so that the experimental energy scale (TeV) does not allow us to access it. With the coordinate transformation $\Theta=R_0 \theta$, the metric becomes
\begin{eqnarray}\label{metric2}
ds^2=a^2(y)\eta_{\mu\nu}dx^{\mu}dx^{\nu}+dy^2+b^2(y)d\Theta^2.
\end{eqnarray}
Here the warp factors $a(y)$ and $b(y)$ are only dependent on the coordinate $y$, and
$\eta_{\mu\nu}=\text{diag}\left(-1,1,1,1\right)$ is the Minkowski metric. With these assumptions, the field equation \eqref{BS-EinsteinEquations} reduces to
\begin{subequations}\label{Einstein's field equations}
\begin{eqnarray}
(\mu, \nu):&&~~~~~~~~\frac{3 a''}{a}+\frac{3 a' b'}{a b}+\frac{3 a'^2}{a^2}+\frac{b''}{b}=-\frac{1}{2}\phi'^{2}-{V_{\Lambda}}(\phi),\\
(y, y):&&~~~~~~~~\frac{4 a' b'}{a b}+\frac{6 a'^2}{a^2}=\frac{1}{2}\phi'^{2}-V_{\Lambda}(\phi),\\
(\theta, \theta):&&~~~~~~~~\frac{4 a''}{a}+\frac{6 a'^2}{a^2}=-\frac{1}{2}\phi'^{2}-V_{\Lambda}(\phi),
\end{eqnarray}
\end{subequations}
and the equation of motion for the scalar field \eqref{eqphi} can be rewritten as
\begin{eqnarray}\label{Scalar field equation}
	\phi''+\left(4 \frac{a'}{a}+\frac{b'}{b}\right)\phi'-\frac{\partial V_\Lambda}{\partial \phi}=0,
\end{eqnarray}
where the prime represents the derivative with respect to $y$. Note that only three equations in eqs. \eqref{Einstein's field equations} and \eqref{Scalar field equation} are independent, while four functions need to be solved. This allows us to make some reasonable assumptions with regard to the scalar field to guarantee our solution {satisfying} the conditions
\begin{eqnarray}\label{conditions}
	a(y)\lvert_{_{ y=0}}=1,\hspace{1.5cm}   a'(y)\lvert_{_{ y=0}}=0,\hspace{1.5cm}
	b(y)\lvert_{_{ y=0}}\neq 0,\hspace{1.5cm}   b'(y)\lvert_{_{ y=0}}=0.
\end{eqnarray}
In this situation, we start with eqs. \eqref{Einstein's field equations} and \eqref{Scalar field equation} to obtain an analytical solution in the following section.


\section{Two braneworld solutions in six-dimensional asymptotically AdS spacetime}
\label{Brane Solution in 6D Spacetime}

In this section, we {look for solutions} satisfying the above conditions \eqref{conditions}.

{

\subsection{Solution 1}

We adopt the following assumption for the scalar potential {$V_{\Lambda}(\phi)$} with
\begin{equation}
V (\phi)=
\frac{k^{2} v^{2}}{2}+\frac{5}{18} k^{2} v^{4}
-\left(k^{2}+\frac{5 k^{2} v^{2}}{8}\right) \phi^{2}+\left(\frac{5 k^{2}}{12}+\frac{k^{2}}{2 v^{2}}\right) \phi^{4}-\frac{5 k^{2} }{72 v^{2}}\phi^{6},
\end{equation}
\begin{equation}
\Lambda=-\frac{5}{18} k^{2} v^{4} .
\end{equation}
Then, from the Einstein equation \eqref{Einstein's field equations},
{the scalar field and the warp factors }are obtained as
\begin{eqnarray}
\phi (y)&=&v \operatorname{tanh}(k y),\\
a(y) &=& b(y)=\mathrm{e}^{-\frac{1}{24} v^{2} \tanh^{2} (k y)} \text{sech}^{\frac{v^{2}}{6}}(k y).
\end{eqnarray}
Here $v$ is a dimensionless parameter and $k$ is a fundamental energy scale with dimension $[k]=L^{-1}$.  $1/k$ stands for the thickness of the brane.

\subsection{Solution 2}
{Another solution can be found if the scalar potential {$V(\phi)$} and the cosmological constant $\Lambda$ are assumed as}
\begin{eqnarray}
V(\phi)&=&
\left(\frac{k^{2}}{2}+\frac{5 k^{2} v^{2}}{24}\right) \phi^{2}
-\left(\frac{5 k^{2}}{24}+\frac{k^{2}}{2 v^{2}}\right) \phi^{4}
+\frac{5 k^{2}}{72 v^{2}}\phi^{6},
\label{Scalarpotential}\\
\Lambda&=&-\frac{5k^2 v^4}{72}.
\end{eqnarray}
Then, {the scalar field and the warp factors} are obtained as
\begin{eqnarray}
    \phi(y)&=&v~\text{sech}(k y), \label{BackgroundScalarField}\\
	a(y) &=& b(y)=\mathrm{e}^{\frac{1}{24} v^2\tanh ^2(k y)}{{\text{sech}^{\frac{v^2}{12}} (k y)}}. \label{warpfactors}
\end{eqnarray}
Unless otherwise specified, the following discussion in this paper is based on the solution 2. For both solutions, the line element has the following form:
\begin{eqnarray}\label{metric2}
ds^2=a^2(y)\eta_{\mu\nu}dx^{\mu}dx^{\nu}+dy^2+a^2(y)d\Theta^2.
\end{eqnarray}

The main difference between solution 1 and solution 2 is that the brane does not split for solution 1 while splitting for solution 2.}

\begin{figure}[H]
\center{
\subfigure[~Scalar potential]{\includegraphics[width=2.4in,height=1.5in]{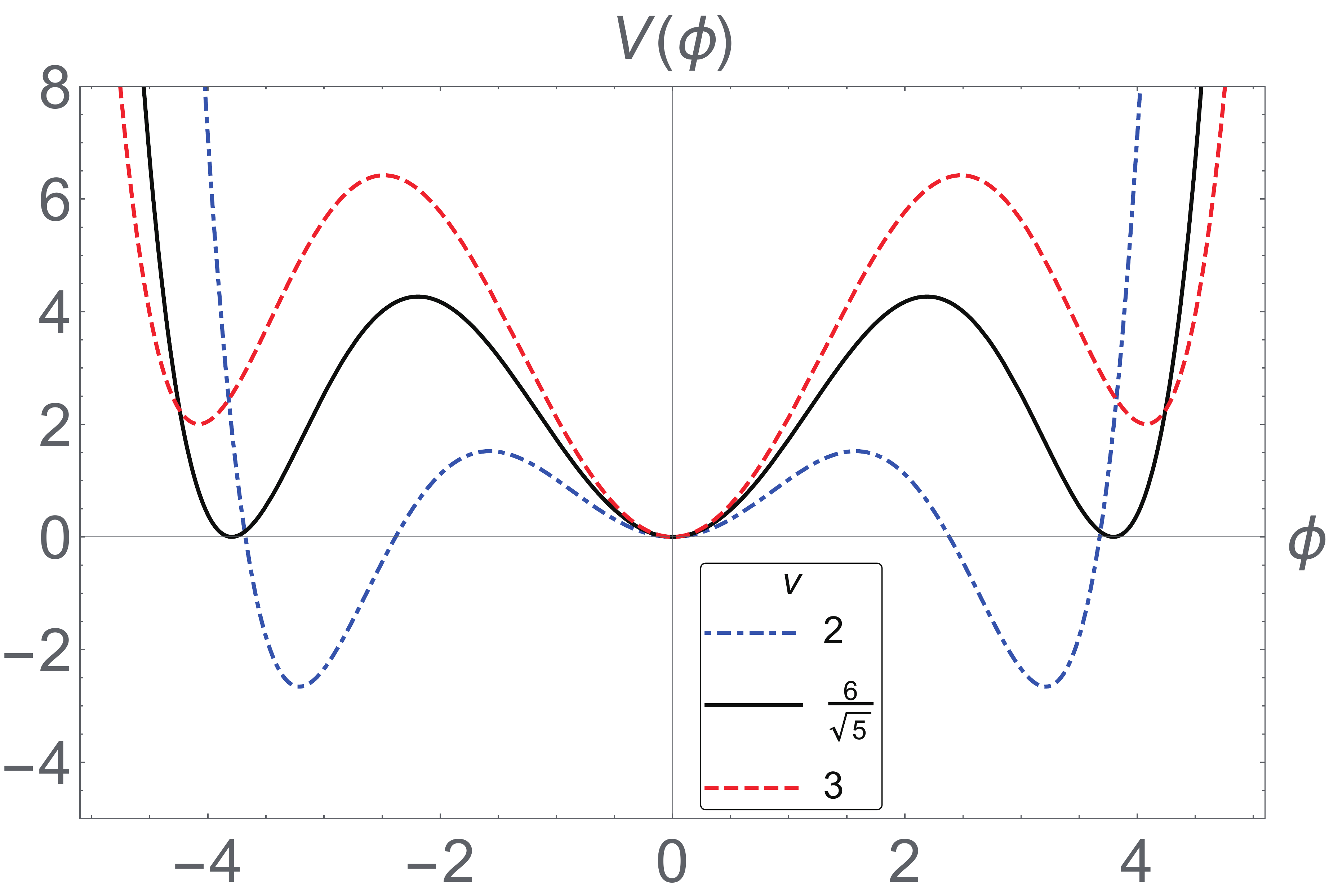}\label{figscalarpotential}}
\quad \quad \quad
\subfigure[~Scalar field]{\includegraphics[width=2.4in,height=1.5in]{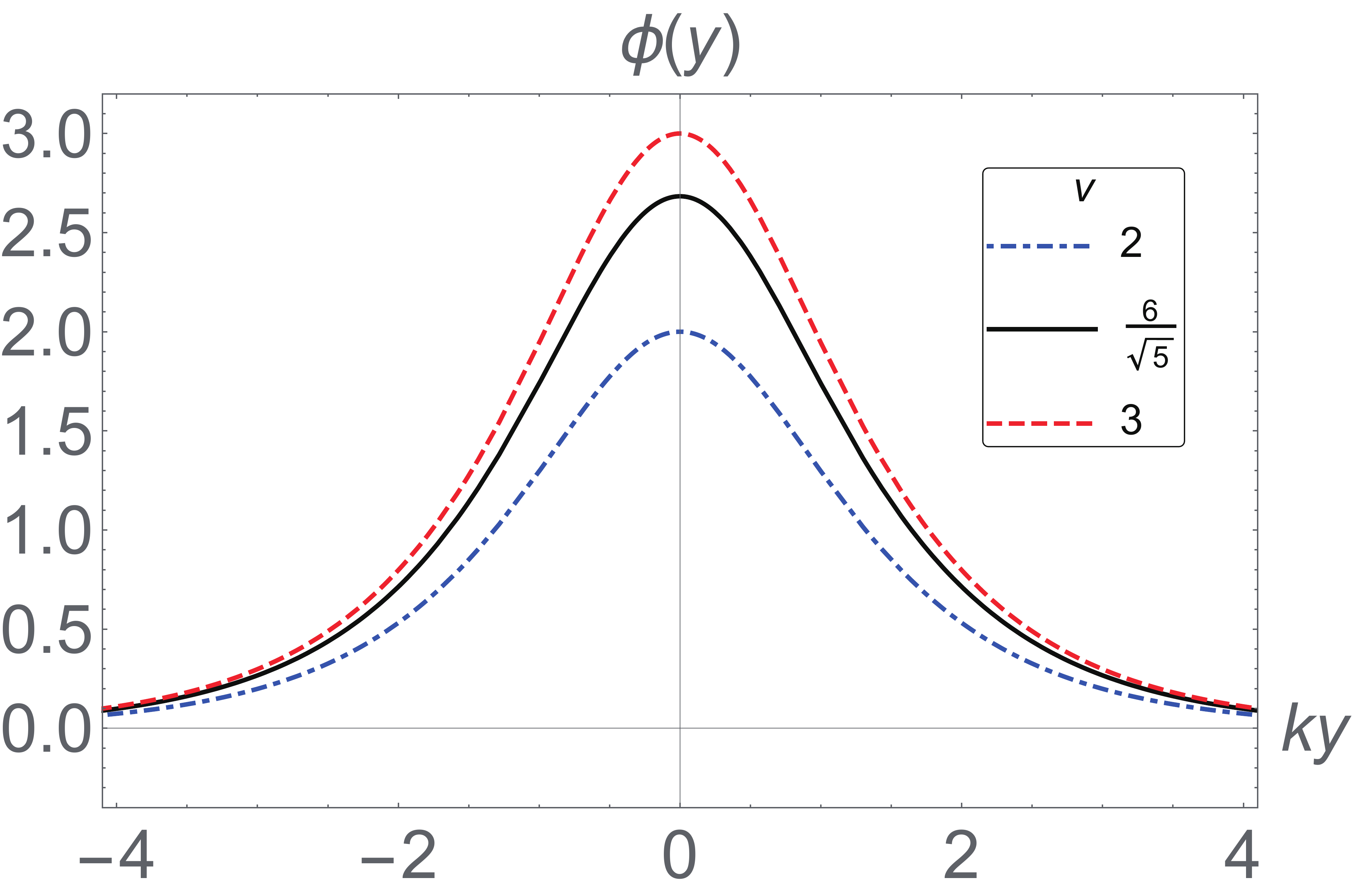}\label{scalarfield}}
}
\caption{Plots of the scalar potential \eqref{Scalarpotential} and background scalar field \eqref{BackgroundScalarField}. The scalar field $\phi(y)\rightarrow 0$ as $y\rightarrow\pm\infty$, which corresponds to the local minimum $V(0)$ of the scalar potential.
}
\label{scalar}
\end{figure}

The scalar potential $V(\phi)$ \eqref{Scalarpotential} and the scalar field $\phi$ \eqref{BackgroundScalarField} are shown in figure \ref{scalar}. Noticing that when $y\rightarrow\pm\infty$, the scalar field $\phi(y)\rightarrow 0$ and the scalar potential $V(\phi)\rightarrow 0$, the contribution to gravity only comes from the cosmological constant. This means that the bulk is an asymptotically AdS$_6$ spacetime. When $v<\frac{6}{\sqrt{5}}$, the AdS vacuum is a false one, and the spacetime may be unstable by considering quantum tunneling. While it corresponds to a true vacuum when $v>\frac{6}{\sqrt{5}}$. When $v=\frac{6}{\sqrt{5}}$, there are three degenerate vacua and the vacuum expectation values are $\left\{0,~\pm  {\frac{6 \sqrt2}{\sqrt5}}\right\}$.

\begin{figure}[H]
\center{
\subfigure[]{\includegraphics[width=2.8in,height=1.6in]{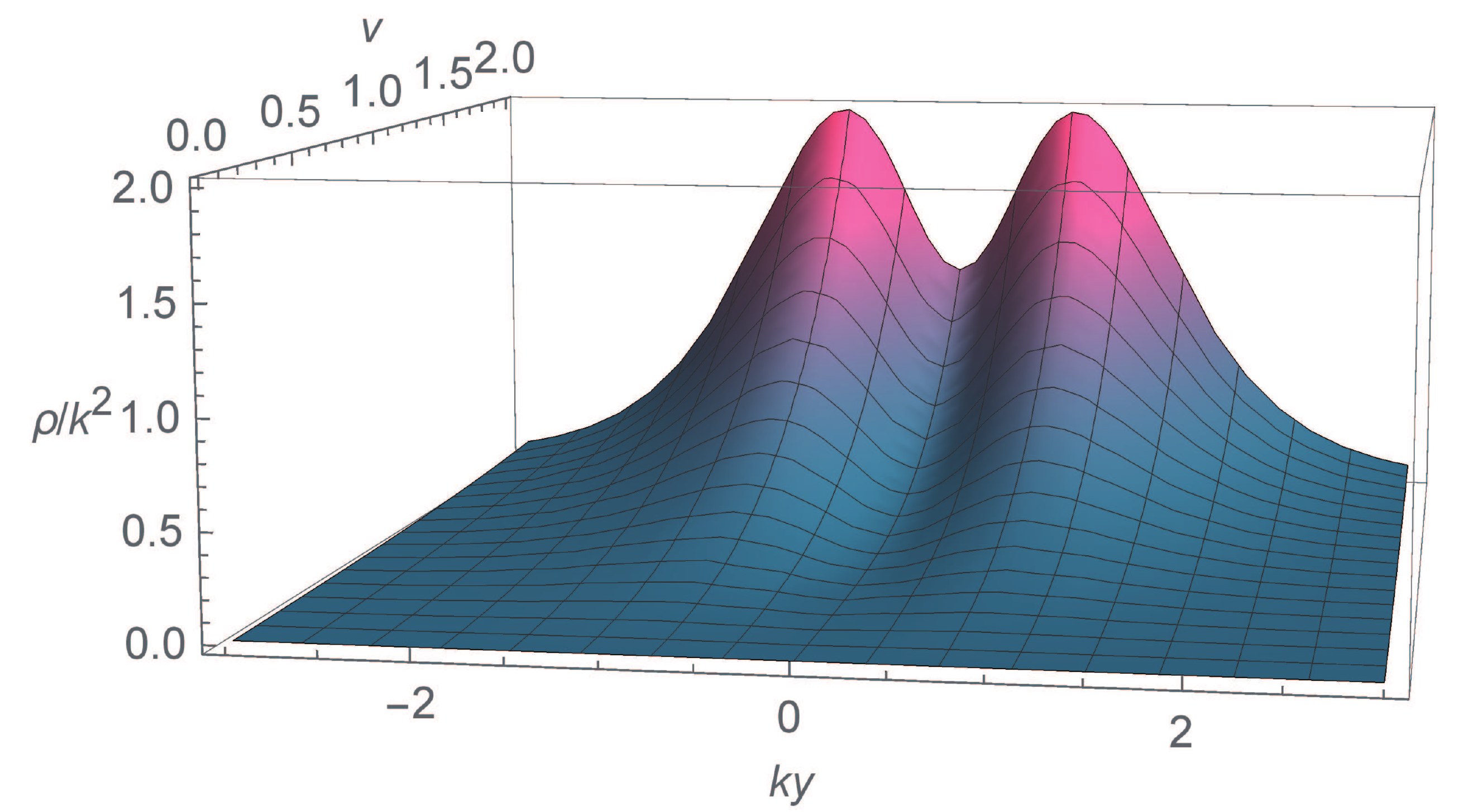}\label{figenergydensity3D}}
\quad \quad \quad
\subfigure[]{\includegraphics[width=2.3in,height=1.5in]{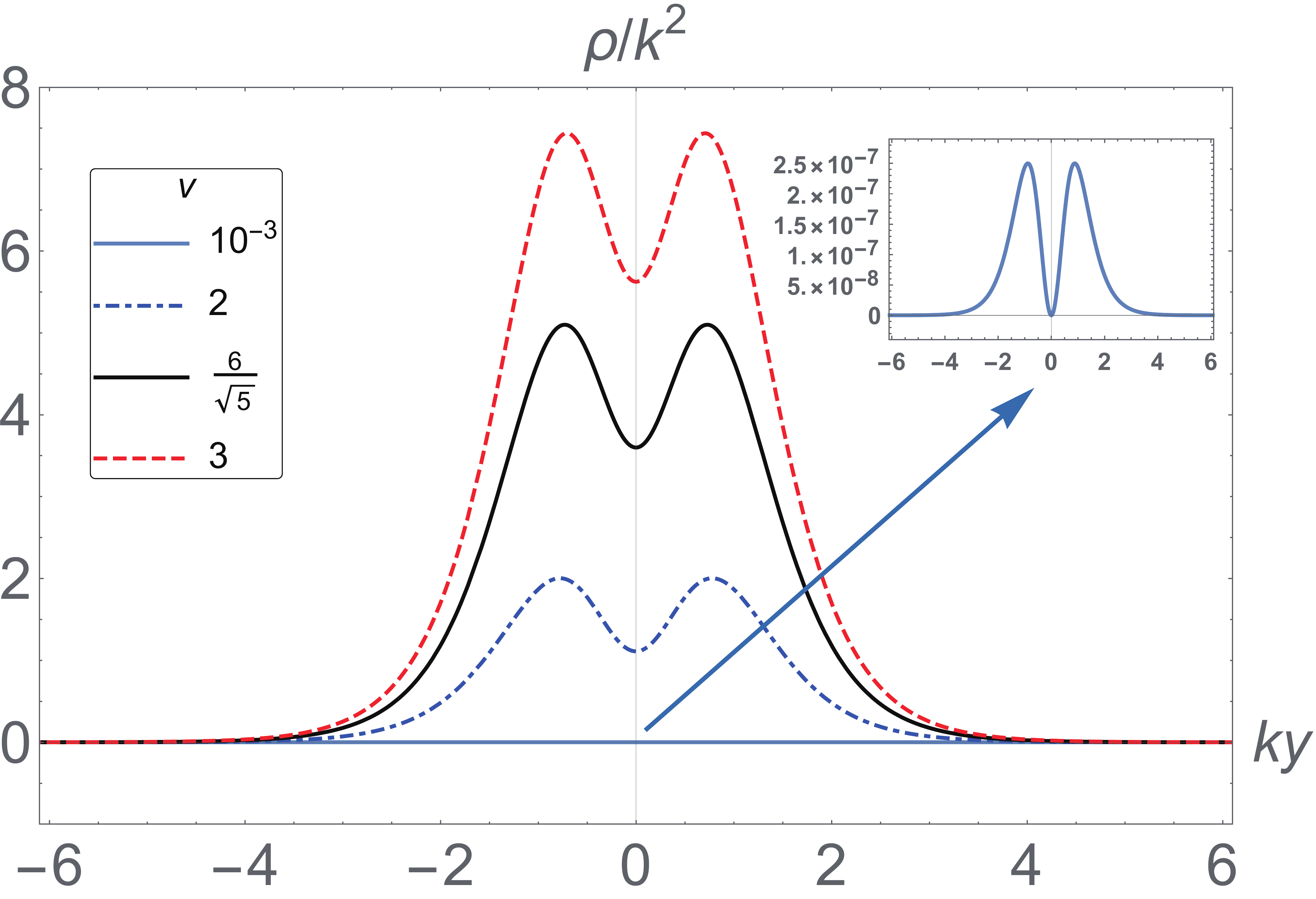}\label{figenergydensity2D}}
}
\caption{The energy density of the background scalar field. Figure \ref{figenergydensity3D} shows the change of the energy density with respect to the parameter $v$, and figure \ref{figenergydensity2D} shows that the brane  splits into two separated sub-branes as $v \rightarrow 0$.
}
\label{figenergydensity}
\end{figure}

For an observer with time-like {6-}veiocity satisfying $g_{MN}U^MU^N=-1$, the energy density of the background scalar field reads
\begin{eqnarray}
	\rho=T_{MN}U^MU^N,
\end{eqnarray}
where the cosmological constant is not included. For a static observer, the energy density $\rho=T^0_{~0}$ is shown in figure \ref{figenergydensity}.

\begin{figure}[H]
\center{
\subfigure[]{\includegraphics[width=2.4in]{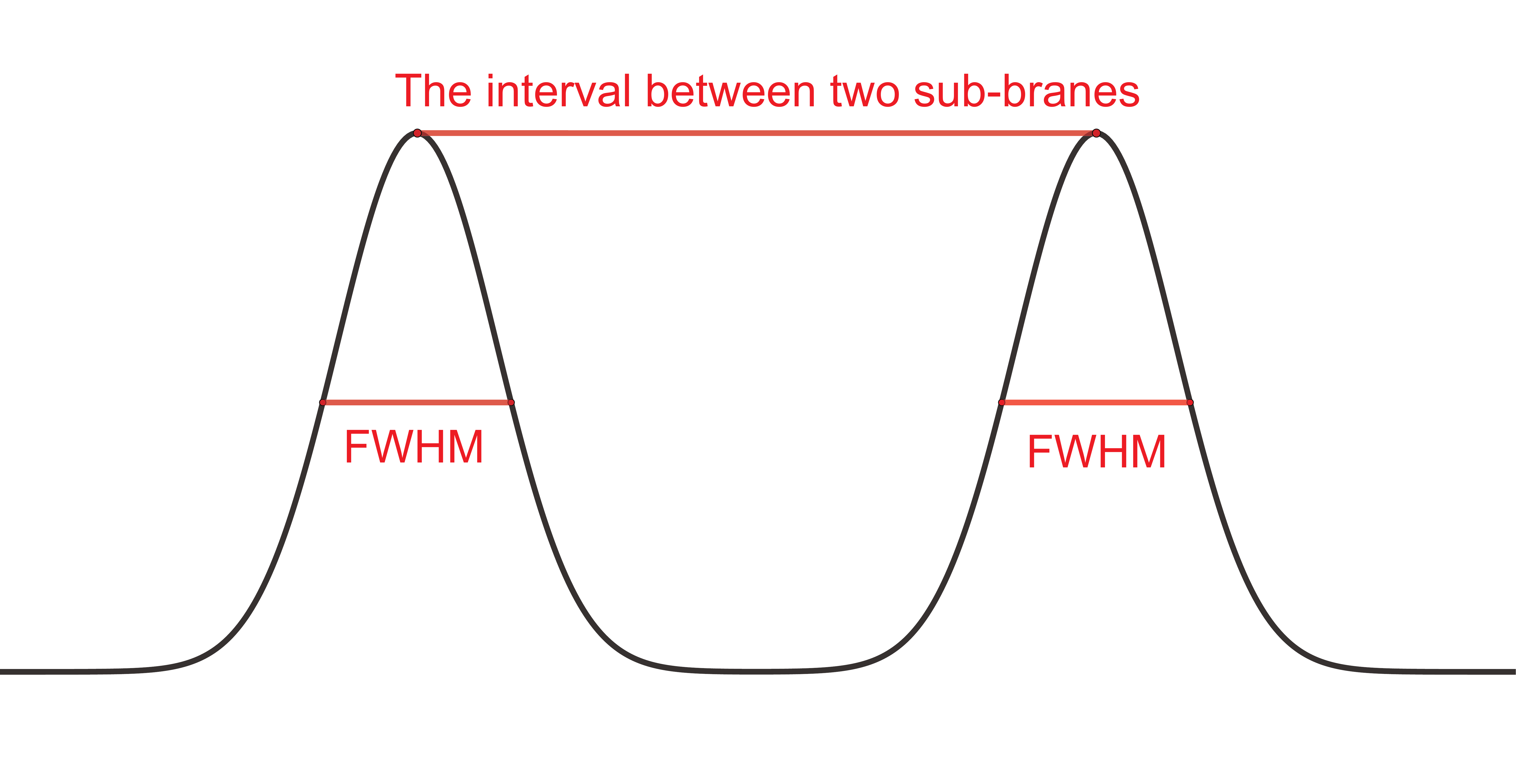}}
}
\caption{The schematic diagram of the degree of separation between two sub-branes.}
\label{schematicDiagram_SeparationDegree}
\end{figure}

The volcano-shaped energy density in figure \ref{figenergydensity} shows that the brane {{has inner structures}}.
By comparing $\rho_0\equiv\rho(0)$ and $\rho_{\text{max}}\equiv \text{Max}[\rho(y)], $ we can define
\begin{equation}\label{splittinglambda}
\lambda(v)=(\rho_{\text{max}}-\rho_0)/\rho_{\text{max}},
\end{equation}
which represents the relative degree of splitting of the brane. Note that this scheme is only suitable for {studying} the degree of brane splitting, it can not well reflect the degree of separation between two sub-branes. When $\lambda(v) > 1/2$,
we can deem that the two sub-branes are separated. In this case, for two symmetric sub-branes,  we can define the full width at half maximum (FWHM) for each sub-brane. Then, we present the following method to measure the degree of separation between the two sub-branes as
\begin{equation} \label{tildelambda}
\tilde{\lambda}(v)
=\frac{\text{Interval between two sub-branes}}{\text{FWHM for each sub-brane}}.
\end{equation}
See figure~\ref{schematicDiagram_SeparationDegree} for the schematic diagram of the above description.

The above definition is to facilitate the generalization to the case of multiple sub-branes. Here we just use the first scheme for our model. We find that $\lambda(v)\rightarrow 1$ as $v \rightarrow 0$ (as shown in figure \ref{figenergydensity2D}), and the degree of splitting $\lambda(v)$ decreases with $v$ (as shown in figure \ref{figenergydensity3D}). The brane is always splitting for any value of $v$, and this result can be obtained from $\rho''(y)>0$ at $y=0$.

\begin{figure}[H]
\center{
\subfigure[~Warp factor]{\includegraphics[width=2.3in,height=1.5in]{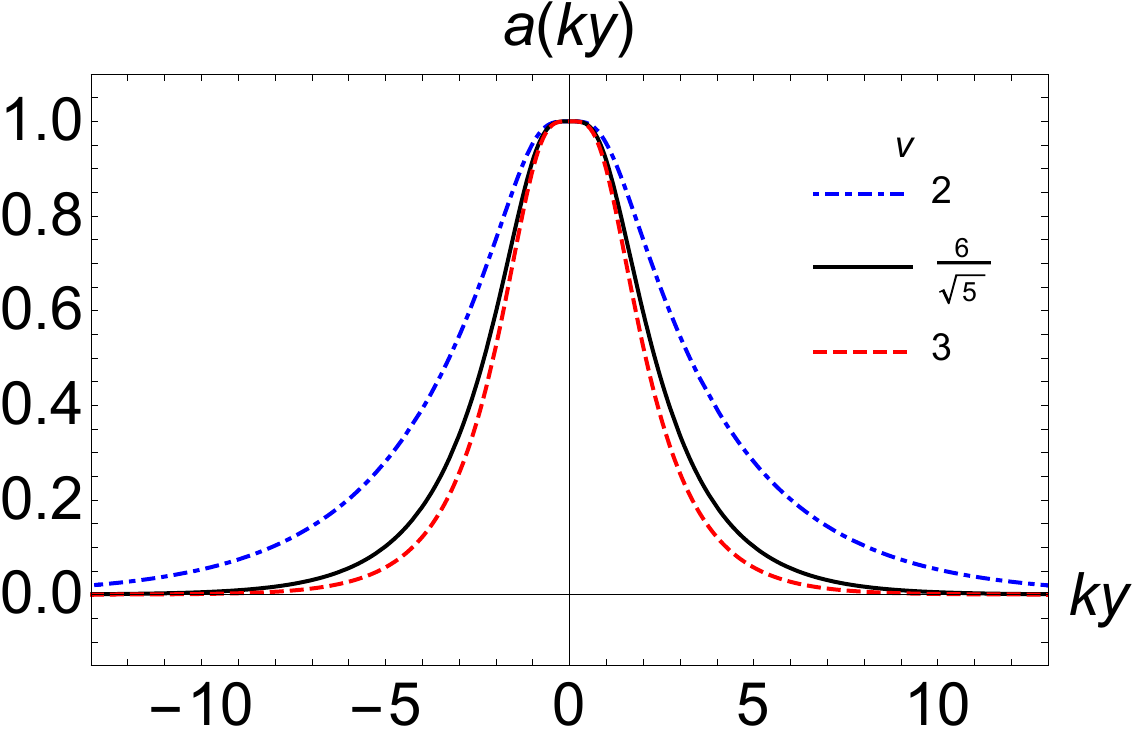}\label{figwarpfactor}}
\hspace{1.2cm}
\subfigure[~Scalar curvature]{\includegraphics[width=2.3in,height=1.5in]{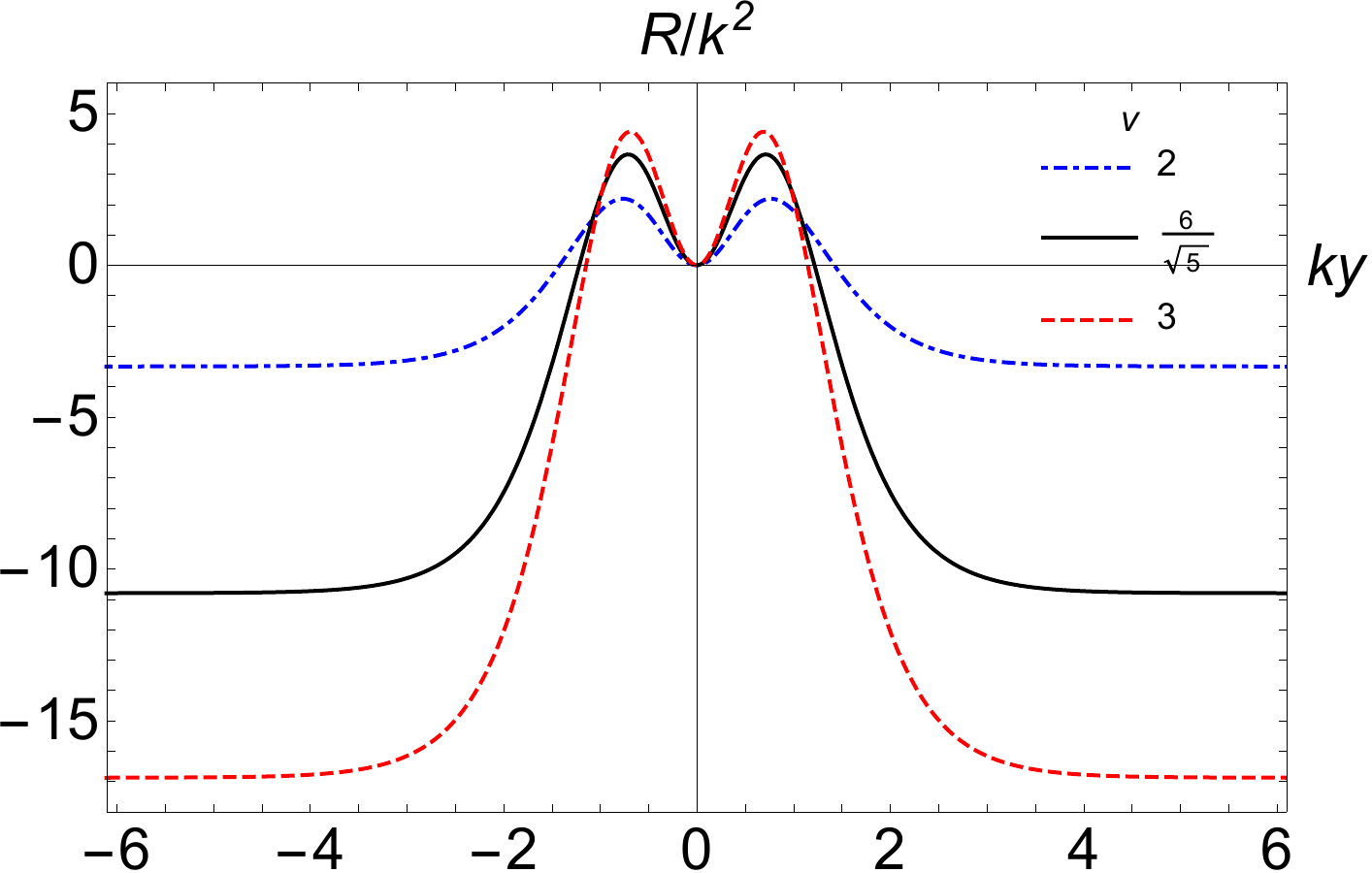}\label{figRscalar}}
}
\caption{Plots of the warp factor \eqref{warpfactors} and the corresponding scalar curvature.
The spacetime is asymptotically $\text{AdS}_6$
at $y\rightarrow \pm \infty$ and Minkowski at $y=0$.}
\end{figure}

\begin{figure}[H]
\center{
\includegraphics[width=2.5in]{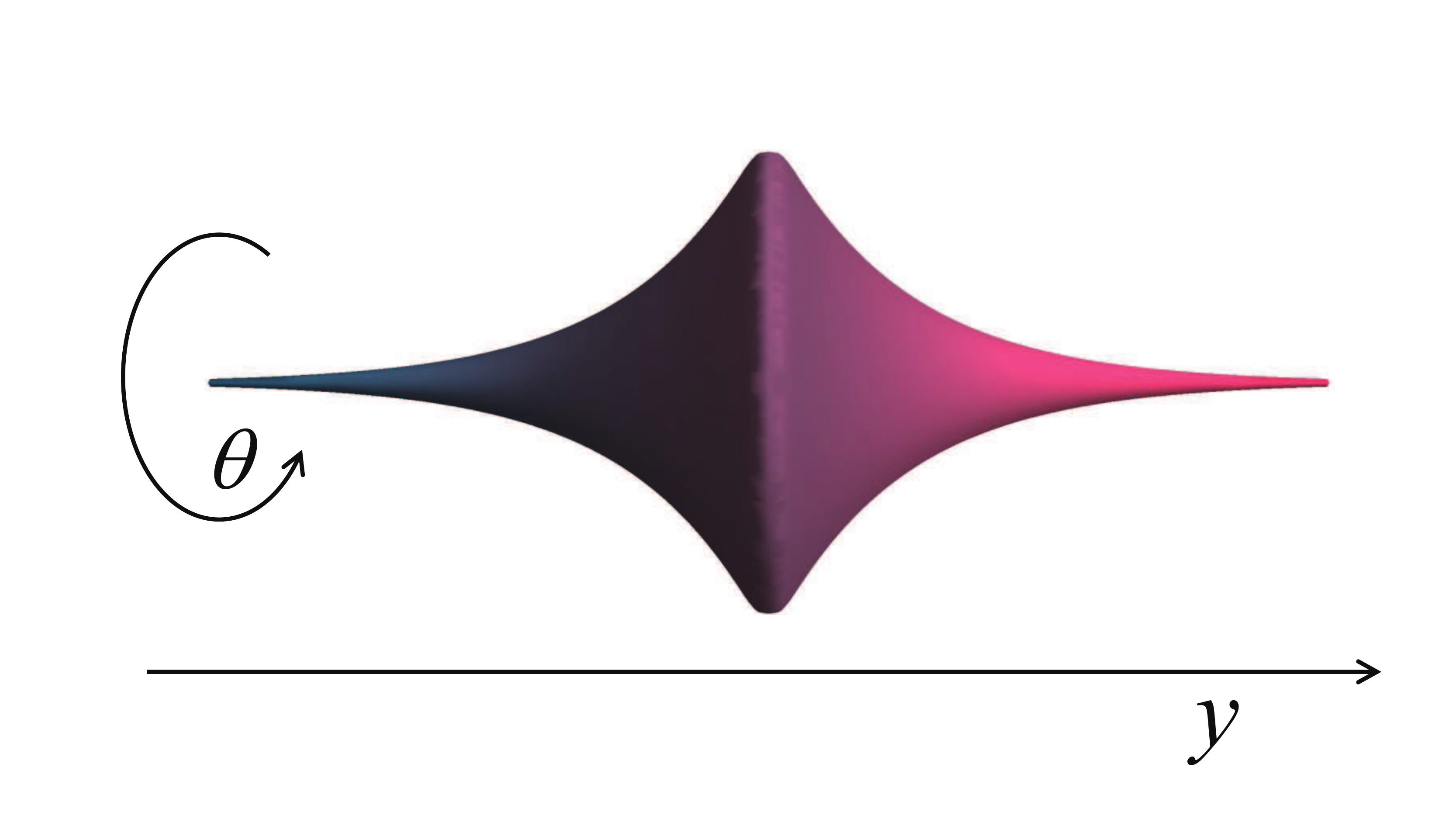}
}
\caption{The profile of the extra dimensions described by the line element
$ds^2_{\text{extra}}=dy^2+b^2(y)R_0^2d\theta^2$. The parameter $v$ is set to $1$.
The compact radius decreases with $|y|$, which {makes} the compact extra dimension {difficult to detect}. With the warp factor $b(y)$ \eqref{warpfactors}, we find that the area of this two-dimensional surface is finite. This means that although there is an infinitely large {transverse space}, the volume of extra dimensions can still be finite. If there is a smooth finite function on the extra-dimensional manifold, its integral with respect to the extra dimensions will be finite.}
\label{geometric images of extra dimension}
\end{figure}

We show the warp factor \eqref{warpfactors}  in figure \ref{figwarpfactor} and also draw the profile of the extra dimensions described by the line element $ds^2_{\text{extra}}=dy^2+b^2(y)R_0^2d\theta^2$ in figure \ref{geometric images of extra dimension}. {The transverse space} contains a noncompact dimension and a compact one. Comparing with a noncompact five-dimensional configuration, the compact sixth dimension will result in an expected effect on localization as shown in section \ref{Localization of Fields}.

\section{Linear tensor perturbations and localization of the graviton}
\label{linear perturbations and localization}

It is well known that the perturbations of the background can be decomposed into three kinds under the four-dimensional Lorentz transformation: the transverse-traceless (TT) tensor mode, the transverse vector modes, and the scalar modes, namely the scalar-vector-tensor decomposition. The three kinds of modes decouple from each other in linear order \cite{Giovannini:2002wi,Giovannini:2002xz,Giovannini:2001xg,Giovannini:2001fh,Giovannini:2002sb}. So each kind of modes can be investigated independently. In this work, we will only investigate the TT tensor mode, which can be written as
\begin{equation}
\delta g_{MN}=a^2(y){
	\left( \begin{array}{cc}
	h_{\mu\nu}(x^\sigma, y, \Theta) & 0_{4\times2} \\
	0_{2\times4} & 0_{2\times2} \\
	\end{array}
	\right )}.
\end{equation}
{Without considering the TT condition}, we show the first-order perturbations of the Ricci tensor
\begin{subequations}
\begin{eqnarray}
\delta R^{}_{\mu\nu}&=&\frac{1}{2} \partial_{\nu}\partial_{\lambda}h^{\lambda}_\mu+\frac{1}{2}\partial_{\mu}\partial_{\lambda}h^{\lambda}_\nu-\frac{1}{2} \square^{(4)}\hmn-\frac{1}{2} \partial_{\mu}\partial_{\nu}h \nonumber\\
&&-aa''\hmn-3a'^2\hmn-2aa'\hmn'-\frac{1}{2} a^2\hmn''-\frac{1}{2} aa'h'\eta_\mn \nonumber\\
&&-\frac{1}{2} \frac{a^2}{b^2} \partial^2_{\Theta}\hmn-\frac{1}{2} \frac{a^2b'}{b}h'_\mn-\frac{aa'b'}{b}\hmn ,\\
\delta R^{}_{\mu y}&=&-\frac{1}{2} \partial_\mu h'+\frac{1}{2} \partial^\lambda h'_{\mu\lambda},\\
\delta R^{}_{\mu \Theta}&=& \frac{1}{2} \partial_{\Theta}\partial_{\lambda}h^\lambda_\mu-\frac{1}{2} \partial_{\Theta}\partial_{\mu}h,\\
\delta R^{}_{{yy}}&=&-\frac{1}{2} h''-\frac{a'}{a}h',\\
\delta R^{}_{{y\Theta}}&=&-\frac{1}{2} \partial_{\Theta}h'+\frac{1}{2}\left(\frac{b'}{b}-\frac{a'}{a}\right)\partial_\Theta h,\\
\delta R^{}_{{\Theta\Theta}}&=&-\frac{1}{2} \partial_{\Theta}^2h-\frac{1}{2} bb'h',
\end{eqnarray}
\end{subequations}
and the first-order perturbation of the scalar curvature
\begin{eqnarray}
\delta R^{}=\frac{1}{a^{2}}\partial_{\nu}\partial_{\lambda}h^{\nu\lambda}-\frac{1}{a^{2}}\square^{(4)}h-\frac{1}{b^2}\partial_{\Theta}^2h
-h''-\left(5\frac{a'}{a}+\frac{b'}{b}\right)h',
\end{eqnarray}
where $\square^{(4)}=\eta^{\mu\nu}\partial_{\mu}\partial_{\nu}$ is the four-dimensional d'Alembert operator and $h=\eta^{\mu\nu}h_{\mu\nu}$. Taking into account the TT condition {$h=0=\partial^{\mu}h_{\mu\nu}$}, the equation for the perturbations $h_{\mu\nu}$ reduces to
\begin{eqnarray}\label{eqofhmunu}
\frac{1}{a^2}\square^{(4)}h_{\mu\nu}+\frac{1}{b^2}{\partial_\Theta^2}h_{\mu\nu}+h''_ {\mu\nu}+\left(4\frac{a'}{a}+\frac{b'}{b}\right)h'_{\mu\nu}=0.
\end{eqnarray}
This equation can also be derived by varying the gravitational action including the quadratic terms of the tensor perturbations. More generally, the above equation can be written as
\begin{eqnarray}\label{KGofgraviton}
{\square^{(6)} \hmn
=0.}
\end{eqnarray}
The corresponding action can be written as
\begin{eqnarray}\label{S2}
S_2 {\sim }\int d^6 x \sqrt{-g} g^{MN}
{\nabla_M} h_{\mu\nu} {\nabla_N} h^{\mu\nu}.
\end{eqnarray}

With the coordinate transformation
\begin{eqnarray}
dz=\frac{dy}{a(y)},\label{coordinatetrans}
\end{eqnarray}
we can rewrite the perturbation equation \eqref{eqofhmunu} as
\begin{eqnarray}
\square^{(4)}h_{\mu\nu}+\frac{a^2}{b^2}{\partial_\Theta}^2h_{\mu\nu}+\partial_z^2h_ {\mu\nu}+\left(\frac{\partial_z b}{b}+3\frac{\partial_z a}{a}\right)\partial_z h_{\mu\nu}=0.
\end{eqnarray}
Performing the KK decomposition
\begin{equation}\label{eqforhmunu}
h_{\mu\nu} (x^{\sigma}, z, \Theta) = \sum_{m,n} \hat{h}^{(m)}_{\mu\nu}(x^{\sigma})\varphi_{(m,n)}(z) e^{i l_n \Theta},
\end{equation}
where $e^{i l_n \Theta}$ is the compact dimensional part, we obtain the Klein-Gordon (KG) equation for the four-dimensional part $\hat{h}^{(m)}_{\mu\nu}(x^{{\sigma}})$:
\begin{equation}\label{4D part KG hmunu}
\left(\square^{(4)}-m^2\right)\hat{h}^{(m)}_{\mu\nu}(x^{{\sigma}})=0,
\end{equation}
and the equation for the noncompact extra-dimensional part $\varphi_{mn}(z)$:
\begin{equation}\label{eqphin}
\partial_z^2\varphi_{(m,n)}+\left(\frac{\partial_z b}{b}+3\frac{\partial_z a}{a}\right)\partial_z \varphi_{(m,n)}+\left(m^2-\frac{a^2}{ b^2}l_n^2\right)\varphi_{(m,n)}=0.
\end{equation}
Here $m$ can be interpreted as the effective mass of the KK graviton, and the azimuthal number $l_n=n/R_0$, where $n$ is an integer due to the periodic boundary condition.

\begin{figure}[H]
\center{
\subfigure[~The effective potential of the graviton]{\includegraphics[width=2.3in,height=1.5in]{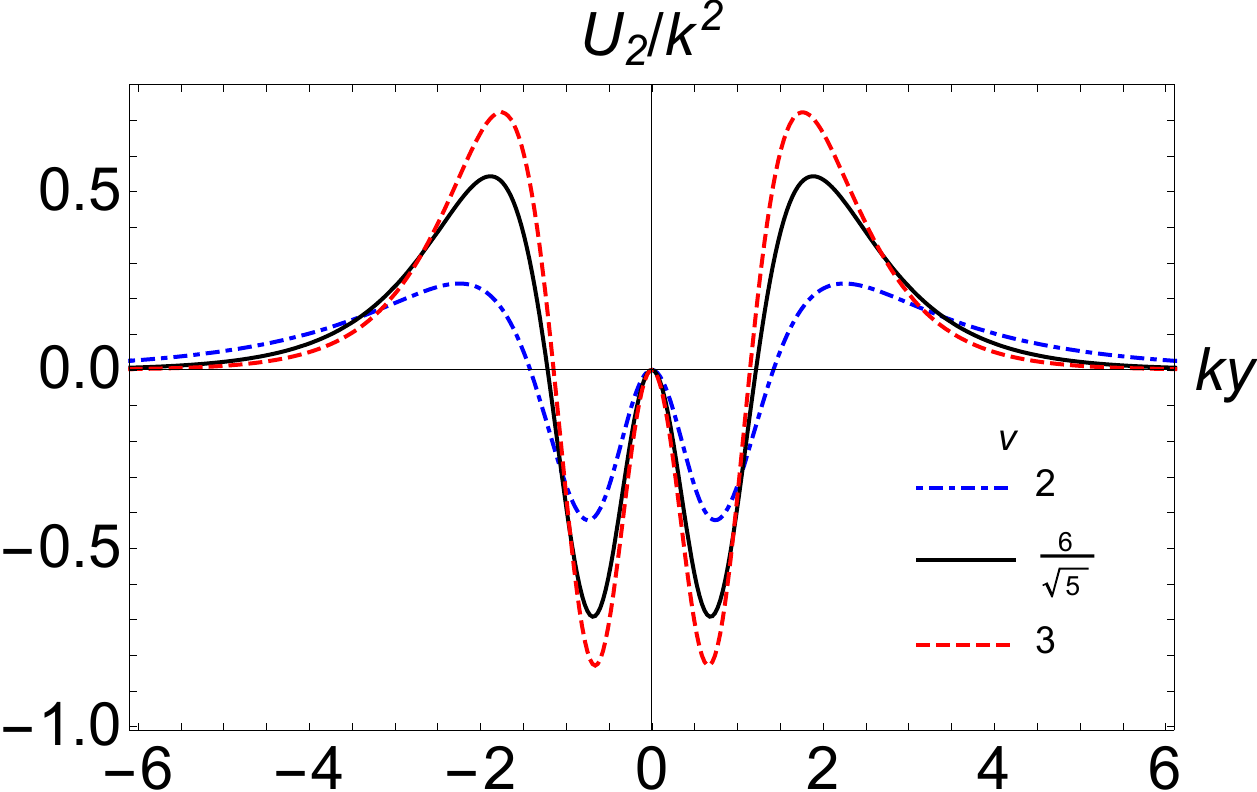}\label{U2}}
\hspace{1.5cm}
\subfigure[~The wave function of the ground state with the eigenvalue $m_n^2-l_n^2=0$]{\includegraphics[width=2.2in,height=1.5in]{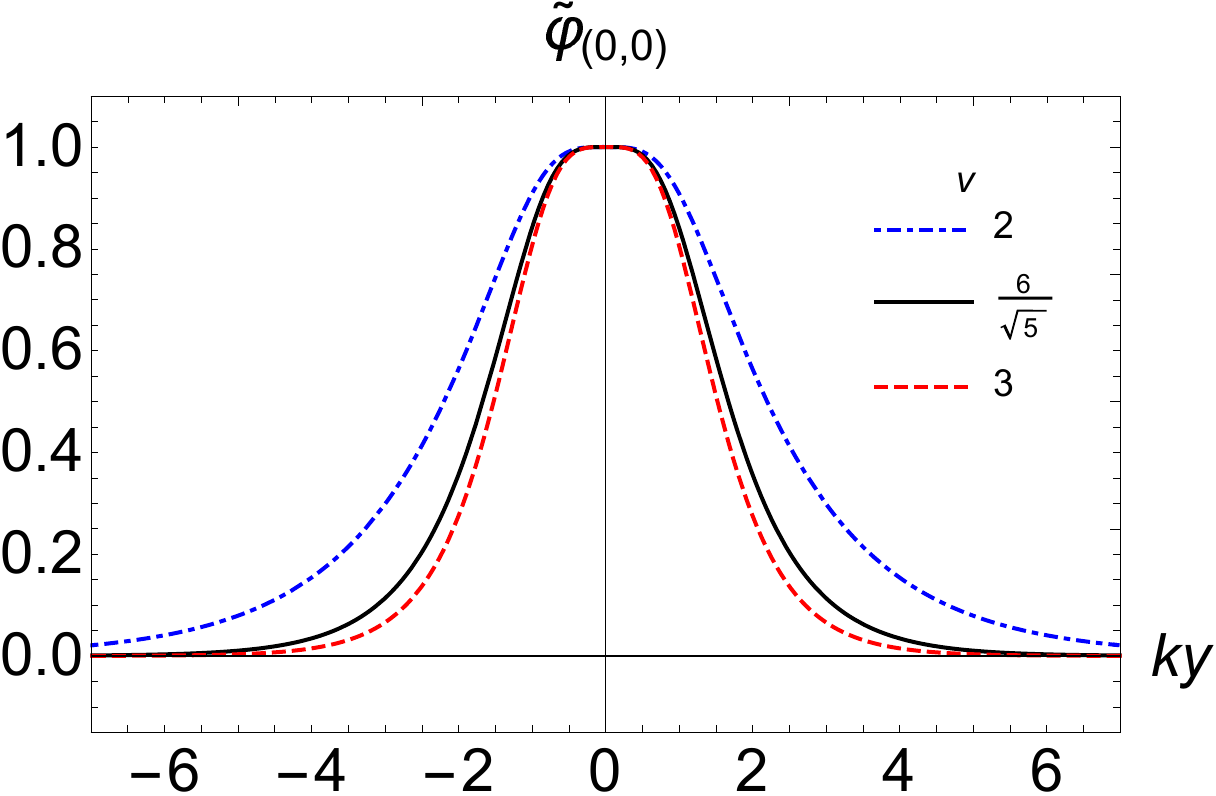}\label{phi0}}
}
\caption{The effective potential and the wave function of the ground state of Eq. \eqref{schrdingerlike}.
}\label{U2_phi0}
\end{figure}

With the transformation
\begin{eqnarray}
\varphi_{(m,n)}=a^{-\frac{3}{2}}b^{-\frac{1}{2}}\tilde{\varphi}_{(m,n)},
\end{eqnarray}
{Eq.} (\ref{eqphin}) can be rewritten as a Schr\"{o}dinger-like equation
\begin{eqnarray}
\left[-\partial_z^2+U_2(z)\right]\tilde{\varphi}_{(m,n)}(z)=\left(m^2-\frac{a^2}{b^2}l_n^2\right)\tilde{\varphi}_{(m,n)}(z),
\label{schrdingerlike}
\end{eqnarray}
where the effective potential $U_2(z)$ has the form
\begin{eqnarray}
U_2(z)=\frac{3}{4}\frac{(\partial_z a)^2}{a^2}-\frac{1}{4}\frac{(\partial_z b)^2}{b^2}+\frac{6}{4}\frac{\partial_z a}{a}\frac{\partial_z b}{b}+\frac{3}{2}\frac{\partial_z^2 a}{a}+\frac{1}{2}\frac{\partial_z^2 b}{b}.
\label{Wrhoab}
\end{eqnarray}
With the braneworld solution \eqref{warpfactors}, the effective potential reduces to
\begin{eqnarray}\label{U2ep}
U_2(z)=2\frac{(\partial_z a)^2}{a^2}+2\frac{\partial_z^2 a}{a}.
\label{Wrhoa}
\end{eqnarray}
As shown in figure \ref{U2}, we can find that $U_2(z)\rightarrow 0$ when $z\rightarrow\infty$. This means that the KK modes with $m^2-{l_n^2}>0$ are free states.

Eq. \eqref{schrdingerlike} can be factorized as
\begin{eqnarray}\label{superschrdingerlike}
\left(-\partial_z-2\frac{\partial_z a}{a}\right)\left(\partial_z-2\frac{\partial_z a}{a}\right)\tilde{\varphi}_{(m,n)}(z)=\left(m^2-l_n^2\right)\tilde{\varphi}_{(m,n)}(z).
\label{factorized}
\end{eqnarray}
We give a simple discussion for several cases. For convenience, we adopt $m_{n}=|l_n|=|n|/R_0$ to label some special values of $m$.
\begin{itemize}
  \item It is clear that Eq. \eqref{superschrdingerlike} has the form as $A^{\dagger}A\tilde{\varphi}_{(m,n)}(z)=(m^2-l_n^2)\tilde{\varphi}_{(m,n)}(z)$, where $A=\partial_z-2(\partial_z a)/a$. According to the Sturm-Liouville theorem, the lowest energy state of the Schr\"{o}dinger-like equation has no zero point in its domain of definition. The eigenvalue  corresponding to the lowest energy state is  $m_n^2-l_n^2 = 0$. {This indicates that} eigenvalues are nonnegative ($m^2-l_n^2 \geqslant 0$). Therefore{,} the system is {spatially stable} under the tensor perturbations in {the linear} order.
  \item To avoid tachyons and to ensure that the system {is time-stable}, $m^2 \geqslant 0$ is required from the KG equation \eqref{4D part KG hmunu}.
  \item The wave function of the ground state ($m_n^2-l_n^2=0$) for Eq. \eqref{superschrdingerlike} is {$\tilde{\varphi}_{(m_n,n)}(z)=C_1a^2(z)$} {(we have ruled out another linearly independent particular solution that does not satisfy the localization condition)}, where $C_1$  is an integration constant.
\end{itemize}
Due to $m^2-l_n^2\geqslant 0$, the state with $m_0=0$ only has the s-wave ($l_0=0$) mode. To obtain the four-dimensional massless graviton, the massless s-wave ($m_0=l_0=0$) mode $\tilde{\varphi}_{(0,0)}(z)$ should be normalizable. The normalization condition for $\tilde{\varphi}_{(0,0)}(z)$ is
\begin{eqnarray}
\int^{+\infty}_{-\infty} |\tilde{\varphi}_{(0,0)}(z)|^2 dz =\int^{+\infty}_{-\infty} |\tilde{\varphi}_{(0,0)}(y)|^2 a^{-1}(y)dy
= C_1^2 \int^{+\infty}_{-\infty}   a^{3}(y) dy =1.
\end{eqnarray}
Here $C_1$ is in fact a {normalization} constant. For the warp factor \eqref{warpfactors}, we have
\begin{eqnarray}\label{normalization condition}
\int_{-\infty}^{+\infty} dy~ a^p(y) < \infty,~\text{when}~p>0.
\end{eqnarray}
In this six-dimensional braneworld model, all the ground states $\tilde{\varphi}_{(m_n,n)}$ with $m_n=|l_n|=|n|/R_0$ can be localized on the brane. This means that there is a discrete spectrum with the effective mass $m_n=|n|/R_0$ in the four-dimensional effective theory.

The four-dimensional effective gravity comes from the contribution of the massless s-wave mode {($m_0=l_0=0$)}, the discrete {modes} of localized four-dimensional massive gravitons ($m_n=|l_n|=|n|/R_0$), and the continuous {modes} ($m^2-l_n^2>0$). Noticing that the massless s-wave mode is localized on the brane, it will lead to the Newtonian potential. Moreover the other KK modes will lead to the corrections to the Newtonian potential.

When the thickness of the brane is small compared with the length corresponding to the fundamental energy scale $M_*$, the gravitational potential between two particles with mass $M_1$ and $M_2$ located at $y=0$ can be written as
\begin{equation}
U(r)\thicksim
 \sum_{n=-\infty}^{\infty}{G_n}\frac{M_1M_2e^{-m_n r}}{r}+\frac{1}{2\pi R_0M^{4}_{*}}\sum_{n=-\infty}^\infty\int_{|{l_n}|}^\infty dm\,
\frac{M_1M_2e^{-mr}}{r}|\bar{\varphi}_{(m,n)}(0)|^2\ .
\label{cnl}
\end{equation}
Here the effective gravitational constant can be calculated by the relative probability density of the KK modes at $y=0$ as
\begin{equation}
{G_n} \sim  \frac{1}{2\pi R_0 M^{4}_{*}}|\bar{\varphi}_{(m_n, n)}(0)|^2= \frac{1}{2\pi R_0M^{4}_{*}}\frac{|\tilde{\varphi}_{(m_n, n)}(0)|^2}{\langle\tilde{\varphi}_{(m_n, n)}|\tilde{\varphi}_{(m_n, n)}\rangle},
\end{equation}
where $\bar{\varphi}_{(m_n, n)}(z)={\tilde{\varphi}_{(m_n, n)}(z)}/\sqrt{\langle\tilde{\varphi}_{(m_n, n)}|\tilde{\varphi}_{(m_n, n)}\rangle}$ are the normalized wave functions with $m_n=|l_n|=|n|/R_0$. Note that we have ${G_n}={G_0}={G_\text{N}}$ for any $n$, where ${G_\text{N}}$ is the Newtonian constant. The contribution of the continuous states comes from all possible KK modes with $m^2-l_n^2>0$. The wavefunctions $\bar{\varphi}_m(y)$ of the continuous states are normalized as plane waves, i.e. to unity over a period at $|y| \rightarrow\infty$ \cite{Csaki:2000fc}. The $d$-dimensional plane wave continuum density of states $m^{d-1}$ has been considered here \cite{Csaki:2000fc} (the number of the infinite extra dimensions $d=1$ in this model).

Recall that we have decomposed the solution to the {equation }of perturbations as a sum over Kaluza-Klein modes. For bound states, the mass gap is $1/R_0$. Noticing that
\begin{equation}
\sum_{n=-\infty}^{\infty}{G_n} \frac{M_1M_2e^{-m_n r}}{r}
 ={G_\text{N}}\frac{M_1M_2}{r}\left( 1+ \frac{2}{e^{r/R_0}-1} \right),
\end{equation}
it is easy to obtain the asymptotic behavior at short ($r\ll R_0$) and long ($r \gg R_0$) distances:
\begin{equation}\label{gravitational potential asymptotic behavior}
\sum_{n=-\infty}^{\infty}{G_n} \frac{M_1M_2e^{-m_n r}}{r} \sim \left\{\begin{array}{ll}
\frac{R_0}{r^{2}}, &\text{for $r\ll R_0$,}  \\
\\
\frac{1}{r}, & \text{for $r\gg R_0$.}
\end{array}\right.
\end{equation}

At distances shorter than the compactification radius $R_0$, Newtonian potential will be modified \cite{ArkaniHamed:1998rs}. {The experimental tests of the gravitational force are usually parametrized by the modified potential} \cite{Kapner:2006si,Kribs:2006mq,Cheng:2010pt}
\begin{equation}
U(r)\sim G \frac{m_{1} m_{2}}{r}[1+\alpha \exp (-r / \lambda)]. \label{Ur}
\end{equation}
From Eq. \eqref{cnl}, considering the contribution of the KK modes with $m_0=0$ and $m_1=1/R_0$, it is easy to get the parameters as $\alpha=2$, $\lambda=R_0$. The inverse-square law has been experimentally verified at sub-millimeter distances \cite{Kribs:2006mq,Kapner:2006si,Adelberger:2003zx,Hoyle:2004cw,Cheng:2010pt,Yang:2012zzb,Ke:2020nbb}, which {imposes} constraints on our model that $R_0$ should be less than sub-millimeter.

{From Eq. \eqref{gravitational potential asymptotic behavior}}, it has been shown that the massive KK modes are suppressed at long distances, or equivalently at low energies. Hence, it makes sense to write down an action which only includes the massless mode with $n=0$. Plugging the massless mode $\bar{\varphi}_{(0,0)}(z)$ into the action \eqref{S2}, we obtain the low energy effective action
\begin{equation}
S_{\text{eff}}=M^{4}_{*} \pi R_0 {\int \bar{\varphi}_{(0,0)}^2(z) dz}  \int d^{4} x \left[-\frac{1}{{4}}\eta^{\mu\nu}\partial_{\mu} \hat{h}_{\alpha\beta}^{(0)}(x^{\sigma})\partial_{\nu}\hat{h}^{(0)\alpha\beta}(x^{\sigma})\right].
\end{equation}
It can also be viewed as a low energy approximation of general relativity in four-dimensional spacetime and the Newtonian potential can be obtained by this action.


\section{Localization of scalar fields and vector fields}
\label{Localization of Fields}

In this section, we investigate the localization of {the bulk massless scalar field and the free $U(1)$ gauge field}, and expect to obtain a
four-dimensional effective description.

\subsection{Scalar fields}

{The action for a massless scalar field in six-dimensional spacetime reads
\begin{equation} \label{SF-action}
{S_0} =  - \frac{1}{2}\int {{d^6}x\sqrt { - g}~ {g^{MN}}{\partial _M}\Phi^\ast {\partial _N}\Phi },
\end{equation}
and the corresponding equation of motion is
\begin{equation}\label{SF-KGEquation1}
\square^{(6)}\Phi = 0.
\end{equation}
With the metric \eqref{metric2} and performing the KK decomposition
\begin{equation}\label{KK decomposition}
\Phi (x^{\sigma}, y, \Theta) = \sum_{m,n} \phi^{(m,n)}(x^{\sigma})\varphi_{(m,n)}(y) e^{i l_n \Theta},
\end{equation}
the action \eqref{SF-action} can be written as (considering the different KK modes are orthogonal to each other)
\begin{eqnarray} \label{SF-action2}
{S_0} &=&  \sum_{m,n} - \frac{1}{2} \Big[
I_{1(m,n)} \int {{d^4}x~ {\eta^{\mu\nu}}{\partial _\mu}\phi^{\ast(m,n)}(x^\sigma) {\partial _\nu}\phi^{(m,n)}}(x^\sigma) \nonumber \\
 && +I_{2(m,n)} \int {d^4}x ~\phi^{\ast(m,n)}(x^\sigma)\phi^{(m,n)}(x^\sigma)
\Big] ,
\end{eqnarray}
{where}
\begin{eqnarray}
I_{1(m,n)}&=&2\pi  \int dy~a^{3}(y) \varphi^*_{(m,n)}(y)\varphi_{(m,n)}(y),\\
I_{2(m,n)}&=&2\pi  \int dy  \Big[a^5(y){\partial _y}\varphi^*_{(m,n)}(y){\partial _y}\varphi_{(m,n)}(y)
 +a^3(y)l_n^2\varphi^*_{(m,n)}(y)\varphi_{(m,n)}(y)\Big].
\end{eqnarray}
{It is easy to see that $I_{2(m,n)}$ corresponds to the mass parameter of a scalar field $\phi^{(m,n)}(x^\sigma)$ in the four-dimensional effective theory.
The localization condition of the zero mode requires}
\begin{eqnarray}\label{localization condition}
I_{1(0,0)}<\infty.
\end{eqnarray}
{The solution of the zero mode ($I_{2(0,0)}=0$) is $\varphi_{(0,0)}(y)=C_1$ (satisfying the field equation and the localization condition).} This means that the scalar field is homogeneously distributed in the transverse space. But it is still localized. The reason is that the spacetime is {curved} and the warp factor contributes to the integral. The following discussion may help us to understand this. The effect of the curved spacetime can be described equivalently in the following way.

{With the coordinate transformation \eqref{coordinatetrans}, we can rewrite line element \eqref{metric2} with conformal {coordinates}}
\begin{eqnarray}
ds^2=a^2 (z) \left(\eta_{\mu \nu }dx^{\mu }dx^{\nu } +dz^2+d\Theta^2\right).
\end{eqnarray}
{Considering the following field transformation}
\begin{eqnarray}
\varphi_{(m,n)}=a^{-2}(z)\tilde{\varphi}_{(m,n)},
\end{eqnarray}
{we have}
\begin{eqnarray}
I_{1(m,n)}&=&2\pi  \int  dy~a^{3}(y) \varphi^*_{(m,n)}(y) \varphi_{(m,n)}(y)= 2\pi \int dz~\tilde{\varphi}^*_{(m,n)}(z)\tilde{\varphi}_{(m,n)}(z), \nonumber \\ \label{Localization conditions} \\
I_{2(m,n)}&=&2\pi  \int  dy~\left[a^5(y){\partial _y}\varphi^*_{(m,n)}(y){\partial _y}\varphi_{(m,n)}(y)+a^{3}(y)l_n^2\varphi^*_{(m,n)}(y)\varphi_{(m,n)}(y)\right]\nonumber\\
&=&2\pi  \int  dz~a^4(z)~{\partial _z}\left[a^{-2}(z)\tilde{\varphi}^*_{(m,n)}(z)\right]{\partial _z}\left[a^{-2}(z)\tilde{\varphi}_{(m,n)}(z)\right]\nonumber\\
&&+2\pi  l_n^2 \int dz~a^4(z)~ \tilde{\varphi}^*_{(m,n)}(z)\tilde{\varphi}_{(m,n)}(z).
\end{eqnarray}
{From Eq. \eqref{Localization conditions}, we notice that the warp factor $a(z)$ {has been} absorbed in the field $\tilde{\varphi}(z)$. The effect of {curved} spacetime is equivalently represented with the transformation $\varphi_{(m,n)}(y)\rightarrow\tilde{\varphi}_{(m,n)}(z)$ which satisfies the following Schr\"{o}dinger-like equation}
\begin{eqnarray}\label{Schrodinger-like equation}
\left[-\partial_z^2+U_0(z)\right]\tilde{\varphi}_{(m,n)}(z)=m^2 \tilde{\varphi}_{(m,n)}(z),
\label{schrdingerlike}
\end{eqnarray}
{where the effective potential $U_0(z)$ has the form}
\begin{eqnarray}\label{U2ep}
U_0(z)=2\frac{(\partial_z a)^2}{a^2}+2\frac{\partial_z^2 a}{a}.
\label{Wrhoa}
\end{eqnarray}
{The zero mode {can be easily solved as}}
\begin{eqnarray}
\tilde{\varphi}_{(0,0)}(z) = {C_2}a^{2}(z), \label{zeromode}
\end{eqnarray}
{which corresponds to the solution $\varphi_{(0,0)}(y)={C_2}$. It is important to emphasize that $\tilde{\varphi}_{(0,0)}(z)$ is a bound state.}

\begin{figure}[H]
\center{
\subfigure[The effective potential (\ref{Wrhoa}).]{\includegraphics[width=2.3in,height=1.5in]{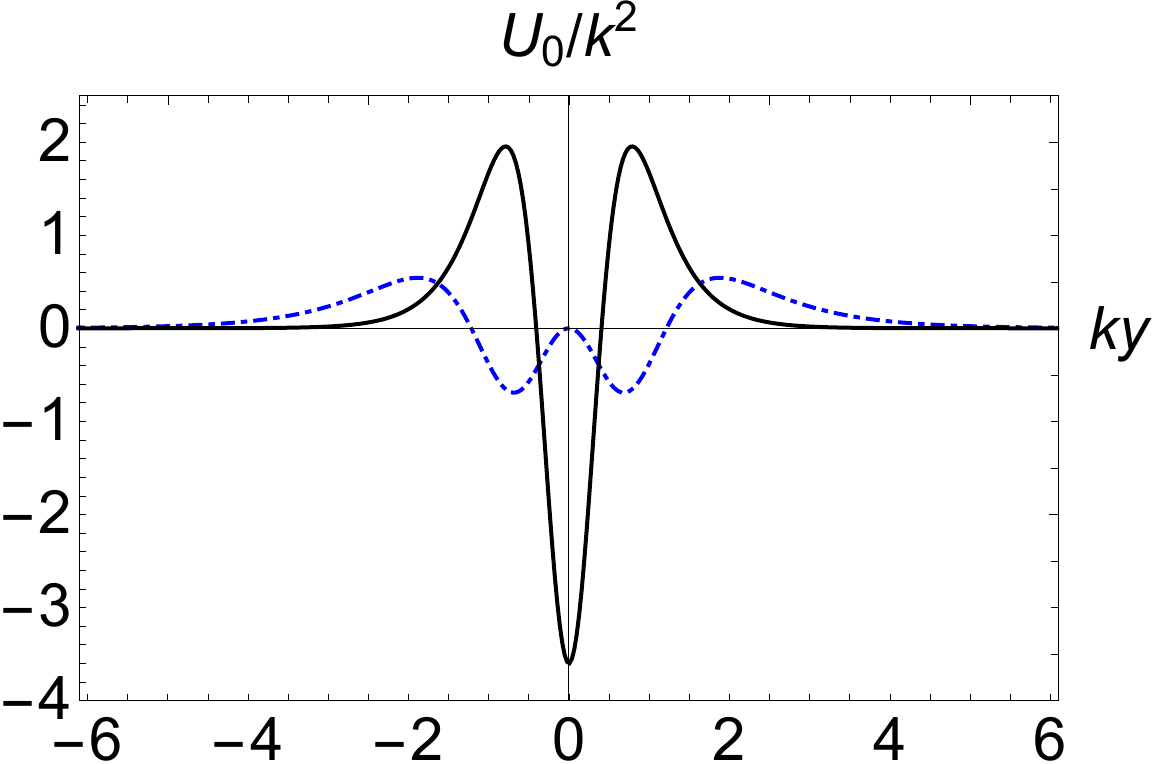}\label{U0}}
\hspace{1.5cm}
\subfigure[~The zero mode (\ref{zeromode}).]{\includegraphics[width=2.2in,height=1.5in]{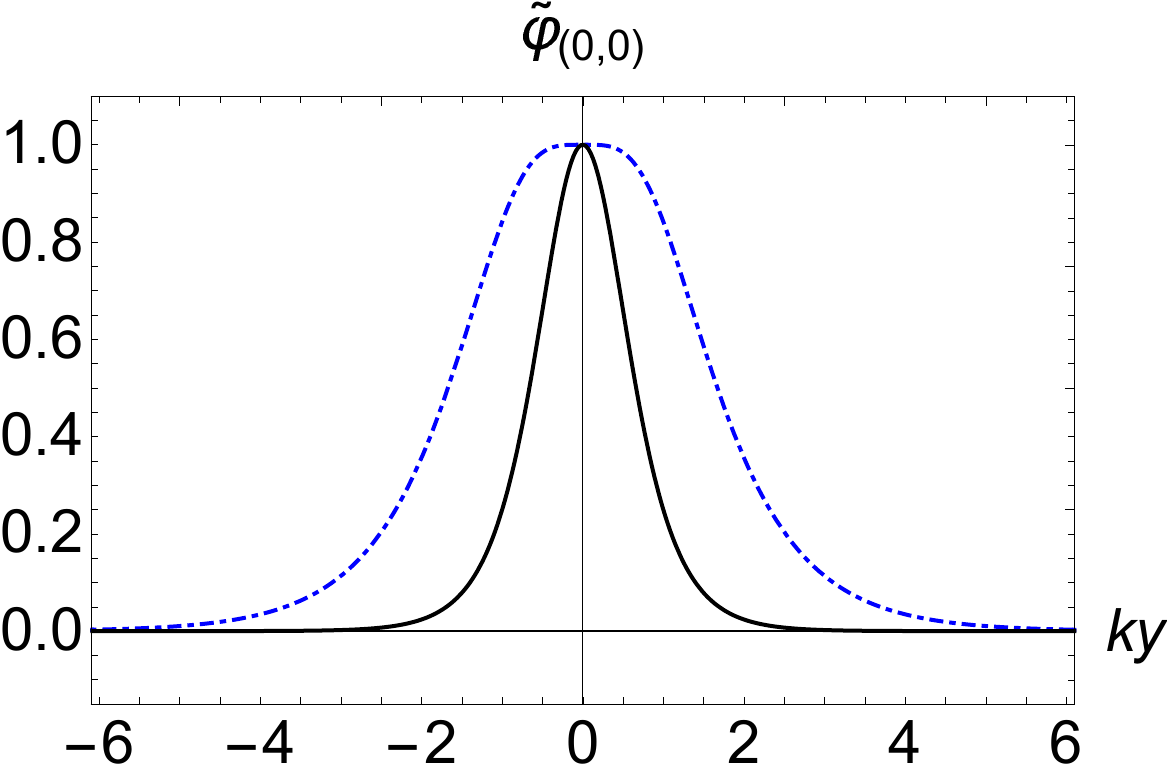}\label{phi0}}
}
\caption{Localization of the scalar field under conformal flat spacetime view. The parameter $v=\frac{6}{\sqrt{5}}$ . The blue dashed (black solid) line is corresponding to solution 2 (solution 1) where the brane is (is not) splitting.
} \label{figure1}
\end{figure}

{Now we can easily consider the localization {{of the scalar field (the analytical method can also be applied to various matter fields)}} from two equivalent points of view.}
\begin{itemize}
  \item {{Physical coordinate view }\\The zero mode ${\varphi}_{(0,0)}(y)={C_2}$  is homogeneously distributed in the transverse space and so it is localized anywhere. {Notice that the
spacetime is curved and there is a minimum coupling between gravity and a scalar field.} The warp factor has a non-trivial contribution to the integral and the localization condition {is satisfied as} $I_{1(0,0)}=2\pi  \int  dy~a^{3}(y) \varphi^*_{(0,0)}(y)\varphi_{(0,0)}(y) < \infty$.}
  \item {{Conformal coordinate view} \\
  Under the conformal coordinates and absorbing the conformal factor $a^2(z)$  into the field function $\tilde{\varphi}_{(m,n)}(z)$, the effect of curved spacetime is reflected by the effective potential (\ref{Wrhoa}) of the Schr\"{o}dinger-like equation. The ground state wave function of the Schr\"{o}dinger-like equation \eqref{schrdingerlike} is a bound state, i.e., $\tilde{\varphi}_{(0,0)}(z)={C_2} a^2(z)$. This zero mode $\tilde{\varphi}_{(0,0)}(z)$ is localized near $z=0$ (see figure \ref{phi0}). The localization condition is satisfied as $I_{1(0,0)}=2\pi  \int  dz  \tilde{\varphi}^*_{(0,0)}(z) \tilde{\varphi}_{(0,0)}(z)< \infty$.}
\end{itemize}
{These two statements are different descriptions of the same effect of gravity on the zero mode, but they are equivalent.}

{With the conformal coordinates view, figure \ref{phi0} shows clearly where the scalar field is localized when the brane splits.
In our six-dimensional model, the scalar zero mode $\tilde{\varphi}_{(0,0)}(z)$ is localized near $z=0$. The zero mode does not split when the brane splits.} By comparing solution 1 and solution 2, it can be found intuitively from figure \ref{phi0} that the scalar zero mode is localized over a wider area along the extra dimension when the brane splits.}

\subsection{Vector fields}

In this subsection, we study the localization of the free $U(1)$ gauge field following the method in ref. \cite{Fu:2018erz}. This method avoids the difficulty of solving the equation of motion of the vector field in a higher-dimensional curved spacetime. The effective action on the brane was derived by the KK decomposition of the vector field, and the consistency condition was obtained. Now we study whether these conditions are satisfied in our model.

We start with the action as
\begin{equation}\label{VectorAction}
S_1 = - \frac{1}{4}\int d^6x\sqrt {-g} F^{M N}F_{M N},
\end{equation}
where
\begin{equation} \label{F}
F_{M N} = \partial _{M} A_{N} - \partial_{N} A_{M}
\end{equation}
is the field strength tensor.

Without loss of generality, we consider the following metric
\beq
ds^2=a^2 (z) \left(\hat{g}_{\mu \nu }dx^{\mu }dx^{\nu } +dz^2+d\Theta^2\right).
\eeq
The quantities with hat ``$\hat{~}$'' represent four-dimensional quantities whose indexes are raised by $\hat{g}^{\mu\nu}$. Such metric includes the case described by our solution. We use the notation $\partial^{\mu}\equiv\hat{g}^{\mu\nu}\partial_{\nu}$ in this section.
This {metric }implies that the four-dimensional part and the extra-dimensional part can be separated through the general KK decomposition
\begin{subequations}
\begin{eqnarray}\label{VectorFieldDecomposition}
A_\mu&=&\sum_{m}\hat{X}_{\mu}^{(m)}(x^{\sigma})~W_1^{(m)}(z,\Theta)~a^{-1},\\
A_z&=&\sum_{m}\hat{\zeta}^{(m)}(x^{\sigma})~W_2^{(m)}(z,\Theta)~a^{-1},\\
A_\Theta&=&\sum_{m}\hat{\xi}^{(m)}(x^{\sigma})~W_3^{(m)}(z,\Theta)~a^{-1}.
\end{eqnarray}
\end{subequations}
We reduce the six-dimensional action to a four-dimensional effective one \cite{Fu:2018erz}
\begin{eqnarray}\label{Effective action}
  S_1&=&-\frac{1}{4}\int {d^6 x} \sqrt{-g}\;F^{MN}F_{MN}\nonumber\\
  &=&-\frac{1}{4}\int {d^6 x} \sqrt{-g}\bigg(F^{\mu_1\mu_2}F_{\mu_1\mu_2}+2F^{\mu_1z}F_{\mu_1z}
  +2F^{\mu_1\Theta}F_{\mu_1\Theta}+2F^{\Theta z}F_{\Theta z}\bigg),
  \nonumber\\
   &=&
     -\frac{1}{4}\sum_m\sum_{m'}\int {d^{4}x} \sqrt{-\hat{g}}
     \;\bigg[
            I_{1}^{(mm')}\;\hat{F}^{(m)}_{\mu_1\mu_2}\;\hat{F}^{\mu_1\mu_2(m')}
            +\big(I_{2}^{(mm')}+I_{4}^{(mm')}\big)\;\hat{X}_{\mu_1}^{(m)}\hat{X}^{\mu_1(m')}\nonumber\\
          &&+I_{3}^{(mm')}\;\partial_{\mu_1}\hat{\zeta}^{(m)}\;\partial^{\mu_1}\hat{\zeta}^{(m')}
            -I_{6}^{(mm')}\;\bigg(\partial_{\mu_1}\hat{\zeta}^{(m)}\;\hat{X}^{\mu_1(m')}
                                +\hat{X}_{\mu_1}^{(m)}\;\partial^{\mu_1}\hat{\zeta}^{(m')}\bigg)\nonumber\\
          &&+I_{5}^{(mm')}\;\partial_{\mu_1}\hat{\xi}^{(m)}\;\partial^{\mu_1}\hat{\xi}^{(m')}
            -I_{8}^{(mm')}\;\bigg(\partial_{\mu_1}\hat{\xi}^{(m)}\;\hat{X}^{\mu_1(m')}
                                +\hat{X}_{\mu_1}^{(m)}\;\partial^{\mu_1}\hat{\xi}^{(m')}\bigg)\nonumber\\
          &&+I_{7}^{(mm')}\;\hat{\zeta}^{(m)}\hat{\zeta}^{(m')}
            +I_{9}^{(mm')}\;\hat{\xi}^{(m)}\hat{\xi}^{(m')}
            -\;I_{10}^{(mm')}\;
            \bigg(\hat{\zeta}^{(m)}\hat{\xi}^{(m')}+\hat{\xi}^{(m)}\hat{\zeta}^{(m')}\bigg)
                                 \bigg],\label{effectiveActionI}
\end{eqnarray}
where $\hat{F}^{(m)}_{\mu\nu} = \partial _{\mu} \hat{X}^{(m)}_{\nu} - \partial _{\nu} \hat{X}^{(m)}_{\mu}$ and the constants are given by
\begin{subequations}\label{constantI}
\begin{eqnarray}
&&{I}_{1}^{(mm')}\equiv\int d \Theta d z~ W_{1}^{(m)} W_{1}^{(m')} ,\\
&&{I}_{2}^{(mm')}\equiv2 \int {d} \Theta {d} z~ \partial_{\Theta}\left(W_{1}^{(m)} a^{-1}\right) \partial_{\Theta}\left({W}_{1}^{(m')} a^{-1}\right) a^{2}  ,\\
&&{I}_{3}^{(mm')}\equiv2 \int {d} \Theta {d} z~ W_{2}^{(m)} {W}_{2}^{(m')} ,\\
&&{I}_{4}^{(mm')}\equiv2 \int {d} \Theta {d} z~ \partial_{z}\left(W_{1}^{(m)} a^{-1}\right) \partial_{z}\left(W_{1}^{(m')} a^{-1}\right) a^{2} ,\\
&&{I}_{5}^{(mm')}\equiv2 \int {d\Theta} {d} z~ {W}_{3}^{(m)} {W}_{3}^{(m')}  ,\\
&&{I}_{6}^{(mm')}\equiv2 \int d \Theta d z~ W_{2}^{(m)} \partial_{z}\left(W_{1}^{(m')} a^{-1}\right) a ,\\
&&{I}_{7}^{(mm')}\equiv2 \int {d} \Theta {d} z~ \partial_{\Theta}\left({W}_{2}^{(m)} a^{-1}\right) \partial_{\Theta}\left({W}_{2}^{(m')} a^{-1}\right) a^{2}  ,\\
&&{I}_{8}^{(mm')}\equiv2 \int {d\Theta dz}~ {W}_{3}^{(m)} \partial_{\Theta}\left({W}_{1}^{(m')} a^{-1}\right) a  ,\\
&&{I}_{9}^{(mm')}\equiv2 \int d \Theta d z~ \partial_{z}\left(W_{3}^{(m)} a^{-1}\right) \partial_{z}\left(W_{3}^{(m')} a^{-1}\right) a^{2}  ,\\
&&{I}_{10}^{(mm')}\equiv2 \int {d} \Theta {d} z~ \partial_{z}\left(W_{3}^{(m)} a^{-1}\right) \partial_{\Theta}\left({W}_{2}^{(m')} a^{-1}\right) a^{2}.
\end{eqnarray}
\end{subequations}
By varying the four-dimensional effective action \eqref{Effective action} with respect to $\hat{X}^{(m)}_{\mu}$, $\hat{\zeta}^{(m)}$, and $\hat{\xi}^{(m)}$, we have
\begin{subequations}\label{Eqn of vector by Effective action}
\begin{align}
&\frac{{I}_{1}^{(mm')} }{\sqrt{-\hat{g}}} \partial_{\mu_{1}}\left(\sqrt{-\hat{g}}~\hat{F}^{(m) \mu_{1} \mu_{2}}\right)
-({I}_{2}^{(mm')}+{I}_{4}^{(mm')})\hat{X}^{\mu_{2}(m)}
+{I}_{6}^{(mm')} \partial^{\mu_2} \hat{\zeta}
+{I}_{8}^{(mm')} \partial^{\mu_2}\hat{\xi}=0,\\
&\frac{I_{3}^{(mm')}}{\sqrt{-\hat{g}}}\partial_{\mu_1}
  \left(\sqrt{-\hat{g}}~\partial^{\mu_1}\hat{\zeta}^{(m')}\right)
  -\frac{I_{6}^{(mm')}}{\sqrt{-\hat{g}}}\partial_{\mu_1} \left(\sqrt{-\hat{g}}~\hat{X}^{\mu_1(m')}\right)
  -I_{7}^{(mm')}\hat{\zeta}^{(m')}+I_{10}^{(mm')}\hat{\xi}^{(m')}=0,\\
&\frac{I_{5}^{(mm')}}{\sqrt{-\hat{g}}}\partial_{\mu_1}
  \left(\sqrt{-\hat{g}}~\partial^{\mu_1}\hat{\xi}^{(m')}\right)
  -\frac{I_{8}^{(mm')}}{\sqrt{-\hat{g}}}\partial_{\mu_1} \left(\sqrt{-\hat{g}}~\hat{X}^{\mu_1(m')}\right)
  -I_{9}^{(mm')}\hat{\xi}^{(m')}+I_{10}^{(mm')}\hat{\zeta}^{(m')}=0.\label{effequ2}
\end{align}
\end{subequations}
On the other hand, by varying the six-dimensional action~(\ref{VectorAction}) with respect to $A_{M}$, we obtain the equation of motion
\begin{equation}\label{VectorFieldEquations}
\frac{1}{\sqrt{- g} }\partial _{M}\left( \sqrt{- g} g^{M P}g^{N Q}F_{P Q} \right) = 0.
\end{equation}
Eq. \eqref{VectorFieldEquations} can be written as
\begin{subequations}\label{VectorFieldEqn6D}
\begin{align}
&\frac{1}{\sqrt{-\hat{g}}}\partial_{\mu_1}\left(\sqrt{-\hat{g}}\hat{F}^{\mu_1\mu_2(m)}\right)
+(\lambda_{1} +\lambda_{2}) \hat{X}^{\mu_2(m)}
-\lambda_{4} \partial^{\mu_2} \hat{\zeta}^{(m)}
-\lambda_{3} \partial^{\mu_2} \hat{\xi}^{(m)}=0, \\
&\frac{1}{\sqrt{-\hat{g}}}\partial_{\mu_1}\left(\sqrt{-\hat{g}}~\hat{g}^{\mu_1\mu_2}\partial_{\mu_2}\hat{\zeta}^{(m)}\right)
-\lambda_{5}\frac{1}{\sqrt{-\hat{g}}}\partial_{\mu_1}\left(\sqrt{-\hat{g}}~\hat{g}^{\mu_1\mu_2}\hat{X}_{\mu_2}^{(m)}\right)
+\lambda_{6} \hat{\zeta}^{(m)}
-\lambda_{7} \hat{\xi}^{(m)}=0, \\
&\frac{1}{\sqrt{-\hat{g}}}\partial_{\mu_1}\left(\sqrt{-\hat{g}}~\hat{g}^{\mu_1\mu_2}\partial_{\mu_2}\hat{\xi}^{(m)}\right)
-\lambda_{8}\frac{1}{\sqrt{-\hat{g}}}\partial_{\mu_1}\left(\sqrt{-\hat{g}}~\hat{g}^{\mu_1\mu_2}\hat{X}_{\mu_2}^{(m)}\right)
-{\lambda_{9} \hat{\zeta}^{(m)}
+\lambda_{10} \hat{\xi}^{(m)}=0},
\end{align}
\end{subequations}
where
\begin{eqnarray}\label{lambda}
\lambda_{1} &\equiv& \frac{\partial_{\Theta}\left(a^{2}\left(\partial_{\Theta}\left(W_{1}^{(m)} a^{-1}\right)\right)\right) a^{-1}}{W_{1}^{(m)}},~~~~~
\lambda_{2}\equiv\frac{\partial_{z}\left(a^{2} \partial_{z}\left(W_{1}^{(m)} a^{-1}\right)\right) a^{-1}}{W_{1}^{(m)}},\nn\\
\lambda_{3}&\equiv&\frac{\partial_{\Theta}\left(W_{3}^{(m)} a\right) a^{-1}}{W_{1}^{(m)}},~~~~~~~~~~~~~~~~~~~~~~~
\lambda_{4}\equiv\frac{\partial_{z}\left(W_{2}^{(m)} a\right) a^{-1}}{W_{1}^{(m)}},\nn\\
\lambda_{5}&\equiv&\frac{\partial_{z}\left(W_{1}^{(m)} a^{-1}\right) a}{W_{2}^{(m)}},~~~~~~~~~~~~~~~~~~~~~~~~
\lambda_{6}\equiv\frac{\partial_{\Theta}\left(\partial_{\Theta}\left(W_{2}^{(m)} a^{-1}\right) a^{2}\right) a^{-1}}{W_{2}^{(m)}},\\
\lambda_{7}&\equiv&\frac{\partial_{\Theta}\left(\partial_{z}\left(W_{3}^{(m)} a^{-1}\right) a^{2}\right) a^{-1}}{W_{2}^{(m)}},~~~~~~~~~\,
\lambda_{8}\equiv\frac{\partial_{\Theta}\left(W_{1}^{(m)} a^{-1}\right) a}{W_{3}^{(m)}},\nn\\
\lambda_{9}&\equiv&\frac{\partial_{z}\left(\partial_{\Theta}\left(W_{2}^{(m)} a^{-1}\right) a^{2}\right) a^{-1}}{W_{3}^{(m)}},~~~~~~~~\,
\lambda_{10}\equiv\frac{\partial_{z}\left(\partial_{z}\left(W_{3}^{(m)} a^{-1}\right) a^{2}\right) a^{-1}}{W_{3}^{(m)}}. \nn
\end{eqnarray}
Note that we have not required $\lambda_i$~$(i=1,2,\cdots,10)$ to be constants yet.

Eq. \eqref{Eqn of vector by Effective action} derived from the {four-dimensional effective} action \eqref{Effective action}
needs to be compatible with Eq. \eqref{VectorFieldEqn6D} from the six-dimensional one \eqref{VectorAction},
which leads to the following consistency conditions:
\begin{align}\label{consistency relationship}
&I_{1}^{(mm')}=\delta^{mm'},~~~~~~~~~I_{2}^{(mm')}=-\lambda_1\delta^{mm'},~~~
I_{4}^{(mm')}=-\lambda_2\delta^{mm'},~~~~I_{6}^{(mm')}=-\lambda_4\delta^{mm'},\nn\\
&I_{8}^{(mm')}=-\lambda_3\delta^{mm'},~~~~I_{3}^{(mm')}=\delta^{mm'},~~~~~~~
I_{6}^{(mm')}=\lambda_5\delta^{mm'},~~~~~~~~I_{7}^{(mm')}=-\lambda_6\delta^{mm'},\nn\\
&I_{10}^{(mm')}=-\lambda_7\delta^{mm'},~~~~I_{5}^{(mm')}=\delta^{mm'},~~~~~~~
I_{8}^{(mm')}=\lambda_8\delta^{mm'},~~~~~~~~I_{9}^{(mm')}=\lambda_{10}\delta^{mm'},\nn\\
&I_{10}^{(mm')}=\lambda_9\delta^{mm'}.
\end{align}
Requiring that $\lambda_i~(i=1,2,\cdots,10)$ are finite constants is a necessary condition but {not sufficient} for the consistency conditions \eqref{consistency relationship}. Furthermore, Eq. \eqref{lambda} can be regarded as the equations by separating variables for Eq. \eqref{VectorFieldEqn6D}.

These conditions guarantee the consistency between the fundamental six-dimensional theory and the {four-dimensional effective} one. Starting from the effective action, one should get compatible results with the fundamental six-dimensional theory. On the other hand, these conditions impose constraints on the higher-dimensional model to ensure that the higher-dimensional theory cannot be incompatible to observations.

We further separate variables as
\begin{subequations}
\beqn
W_1^{(m)}(z,\Theta)&=&\sum_{n}w^{(m,n)}_1(z)e^{il_n\Theta},\\
W_2^{(m)}(z,\Theta)&=&\sum_{n}w^{(m,n)}_2(z)e^{il_n\Theta},\\
W_3^{(m)}(z,\Theta)&=&\sum_{n}w^{(m,n)}_3(z)e^{il_n\Theta}.
\eeqn
\end{subequations}
Then, Eq. \eqref{lambda} can be rewritten as
\begin{eqnarray}
\lambda_{1}=-{{l_n}^2},~~~&&
\lambda_{2} {w}_{1}^{(m,n)}=\partial_{{z}}\left({a}^{2} \partial_{{z}}\left({w}_{1}^{(m,n)} {a}^{-1}\right)\right) {a}^{-1},\nn\\
\frac{\lambda_{3}}{{i} {l_n}} w_{1}^{(m,n)}=w_{3}^{(m,n)},~~~&&
\lambda_{4} {w}_{1}^{(m,n)}=\partial_{z}\left({w}_{2}^{(m,n)} {a}\right) {a}^{-1},\nn\\
\lambda_{5} w_{2}^{(m,n)}=\partial_{z}\left(w_{1}^{(m,n)} a^{-1}\right) a,~~~&&
\lambda_{6}=-{{l_n}^2},\\
\lambda_{7} {w}_{2}^{(m,n)}=i{l_n} \partial_{z}\left(w_{3}^{(m,n)} a^{-1}\right)a,~~~&&
\frac{\lambda_{8}}{{i} {l_n}} {w}_{3}^{(m,n)}={w}_{1}^{(m,n)},\nn\\
\lambda_{9} {w}_{3}^{(m,n)}={i} {l_n} \partial_{z}\left({w}_{2}^{(m,n)} {a}\right) {a}^{-1},~~~&&
\lambda_{10} w_{3}^{(m,n)}=\partial_{z}\left(\partial_{z}\left(w_{3}^{(m,n)} a^{-1}\right) a^{2}\right) a^{-1}.\nn
\end{eqnarray}

\begin{figure}[H]
\center{
\subfigure[~The effective potential of the Schr\"{o}dinger-like equation. ]{\includegraphics[width=2.3in,height=1.5in]{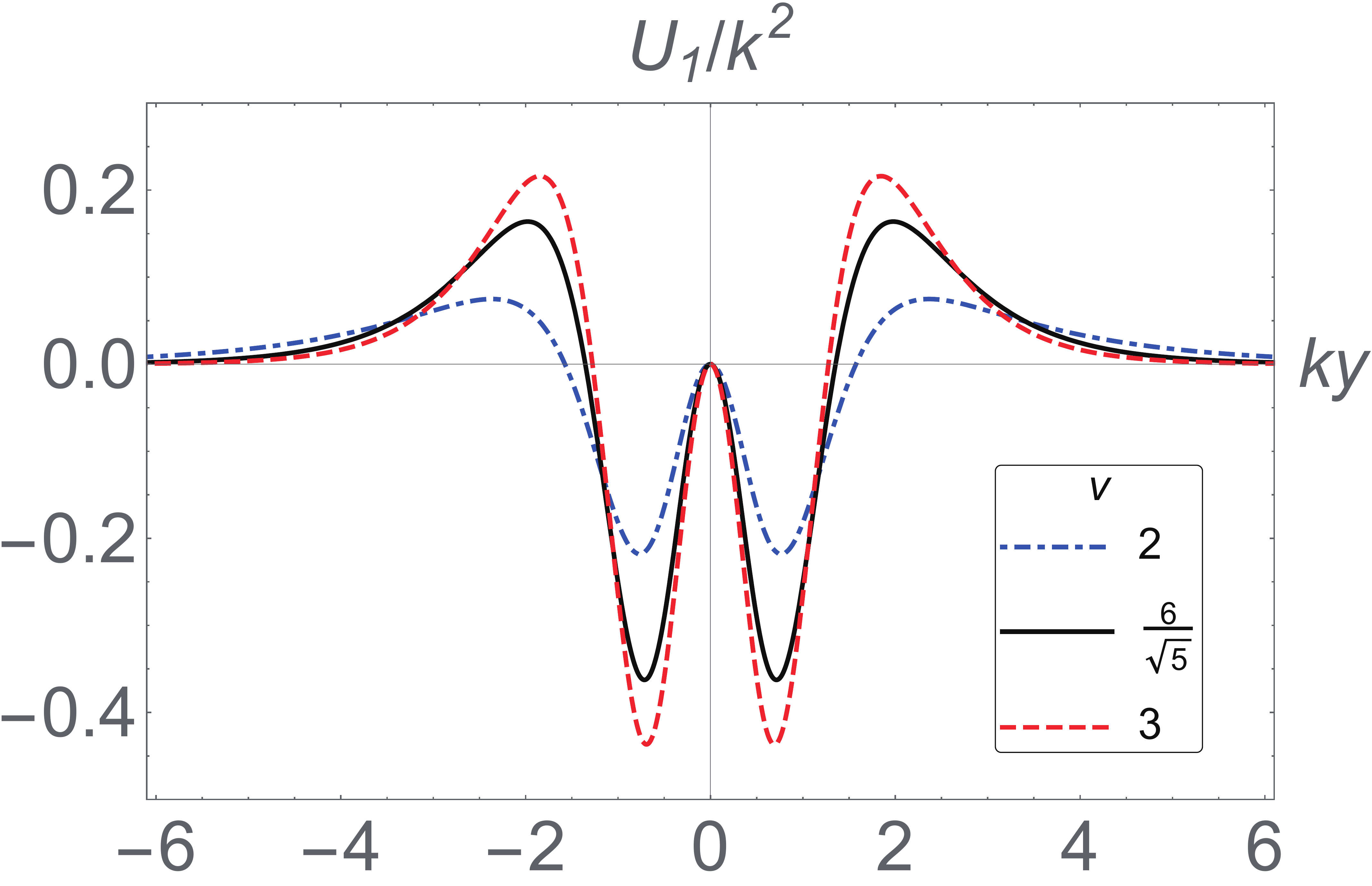}\label{U1}}
\hspace{1.5cm}
\subfigure[~The wave function of the ground state with the eigenvalue $m_n^2-l_n^2=0$.]{\includegraphics[width=2.2in,height=1.5in]{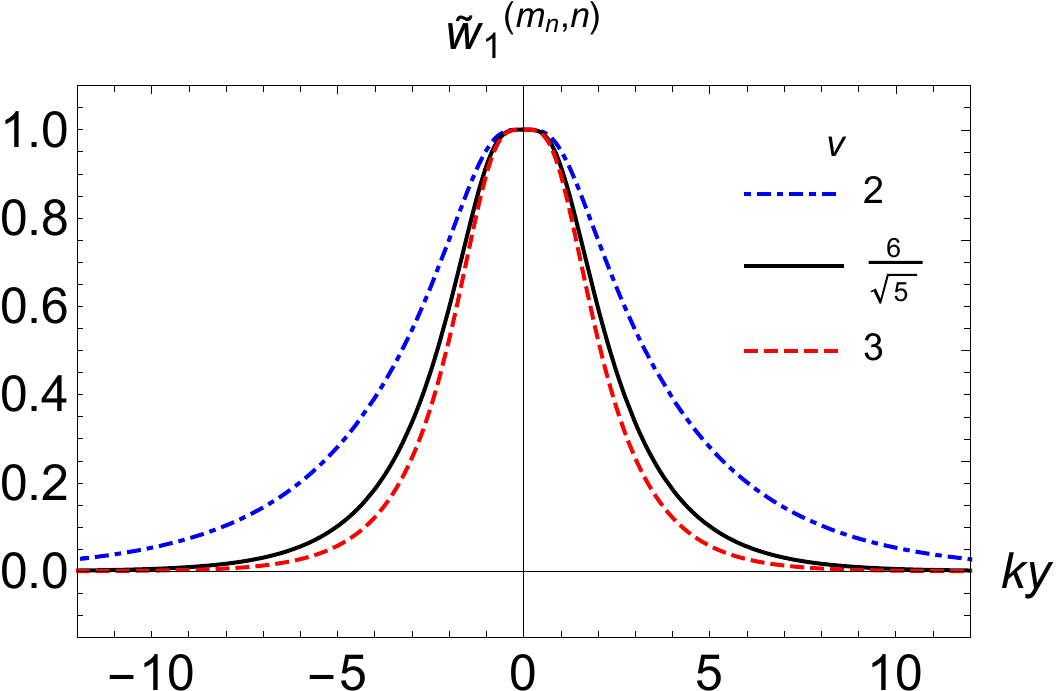}\label{w0}}
}
\caption{The effective potential and the wave function of the ground state of Eq. \eqref{Schrodinger-like equation u1}.
}
\end{figure}

By canonically normalizing the action \eqref{Effective action}, {the four-dimensional effective action can be written as}
\begin{equation}\label{Effective action2}
S=-\frac{1}{4}\sum_m\sum_{m'}\int {d^{4}x} \sqrt{-\hat{g}}
     \;\bigg(
            \;\hat{F}^{(m)}_{\mu_1\mu_2}\;\hat{F}^{\mu_1\mu_2(m')}
            +\frac{I_{2}^{(mm')}+I_{4}^{(mm')}}{I_{1}^{(mm')}}\;\hat{X}_{\mu_1}^{(m)}\hat{X}^{\mu_1(m')}+\cdots\bigg).
\end{equation}
{{Taking into account the consistency conditions \eqref{consistency relationship}, we have}
\begin{eqnarray}\label{58}
  \lambda_1+\lambda_2=-\frac{I_{2}^{(mm')}+I_{4}^{(mm')}}{I_{1}^{(mm')}},
\end{eqnarray}
the effective mass $m$ is given by
\begin{eqnarray}
  m^2=\lambda_1+\lambda_2.
\end{eqnarray}}
Substituting  $\lambda_1$ and $\lambda_2$ defined in \eqref{lambda} into Eq. \eqref{58}, we obtain
\begin{equation}\label{Schrodinger-like equation u1}
\left[ {-\partial_z^2 + U_1(z)} \right]w_1^{(m,n)} =\big(m^2-{l_n^2}\big)w_1^{(m,n)}~,
\end{equation}
where the effective potential $U_1(z)$ is
\begin{eqnarray}\label{U1ep}
U_1(z)=\frac{\partial_z^2 a}{a}
\end{eqnarray}
{which} is shown in figure \ref{U1}. The Schr\"{o}dinger-like equation \eqref{Schrodinger-like equation u1} can be factorized as
\begin{eqnarray}\label{u1 Schrodinger-like equation}
\left(\partial_z+\frac{\partial_z a}{a}\right)\left(-\partial_z+\frac{\partial_z a}{a}\right)w_1^{(m,n)}=\left(m^2-{l_n^2}\right)w_1^{(m,n)}.
\end{eqnarray}
The above equation possesses the form as $B^{\dagger}B W_1^{(m,n)}=(m^2-{l_n^2})W_1^{(m,n)}$ with $B=-\partial_z+(\partial_z a)/a$, which shows that $m^2-{l_n^2} \geqslant 0$ for the warp factor considered in this paper and hence there is no tachyon state.

The equation \eqref{u1 Schrodinger-like equation} has the massless s-wave solution $w_1^{(0,0)}(z)=C_3 a(z)$,
where $C_3$ is a constant. To obtain the four-dimensional Maxwell electromagnetic theory, the $w_1^{(0,0)}(z)$ should be normalizable. According to the consistency conditions \eqref{consistency relationship}, the normalization condition for the massless s-wave mode is
\begin{equation}
I_1^{(00)}=\int^{+\infty}_{-\infty} |w_1^{(0,0)}(z)|^2 dz =\int^{+\infty}_{-\infty} |w_1^{(0,0)}(y)|^2 a^{-1}(y)dy
= C_3^2 \int^{+\infty}_{-\infty} a(y) dy =1.
\end{equation}
From Eq. \eqref{normalization condition}, {we find that} the normalization condition can be guaranteed. Hence the massless s-wave mode can be localized on the brane and the four-dimensional Maxwell electromagnetic theory can be recovered.

In addition, substituting the definitions \eqref{constantI} and \eqref{lambda} into the consistency conditions \eqref{consistency relationship}, we find that the consistency conditions \eqref{consistency relationship} have no contradiction with each other in this model.


\section{Conclusions and discussions}
\label{Conclusions and Outlook}

Using a real scalar field as the dynamic field, we obtained {two analytical solutions} of a smooth thick brane with a compact extra dimension and an infinite one. The bulk is an asymptotically $\text{AdS}_6$ spacetime and is conformally flat. Furthermore, the spacetime is stable under the tensor perturbations in the linear order, and the Newtonian potential on the brane can be recovered. {In fact, our model can be regarded as a six-dimensional extension of the RS-2 model.
Similar to the motivation of the RS-2 model, we achieved the localization of {the graviton} in the case {with an infinite extra dimension and a compact one}. But the free $U(1)$ gauge field is also localized. This is an advantage compared with the five-dimensional thick brane {extension} of the RS-2 model.}

With the background solution, we analyzed the tensor perturbations. Through the KK decomposition of the higher-dimensional graviton,
we found that the massless s-wave mode can be localized on the brane. {Phenomenologically, the Newtonian gravity is recovered on the brane.  The massive KK gravitons will give a correction {to} the Newtonian potential.
The correction includes the contribution of the continuous spectrum (like the RS-2 model) and the discrete one.}
Different from the five-dimensional RS-2 model, there are a series of localized massive modes of the graviton on the brane in our model. The experimental verification of the inverse-square law requires that $R_0$ should be less than sub-millimeter.

{The localization of {the massless} scalar field is similar to that of gravity, and we interpreted the localization picture through two equivalent viewpoints, i.e., the curved spacetime viewpoint and the conformal flat spacetime viewpoint.}
{The splitting of the brane may imply the splitting of the effective potentials of the KK modes of matter fields \cite{Liu:2009ve,Li:2010dy,Zhao:2010mk,Zhang:2016ksq,Sui:2020fty,Yu:2015wma,Zhong:2018fdq}. The calculation in this paper shows that the four-dimensional zero-mass modes of the gravitational field, scalar field and gauge field are still localized near $z=0$, although the effective potentials split for solution 2. {By comparing solution 1, it can be found that the zero modes are localized over a wider area along the extra dimension for solution 2}.} Through the analysis of the supersymmetric partner potentials of the effective potentials, the resonant states of matter fields may occur in solution 2 scenario. These will lead to new physical effects which will help us to test this model. Further research could be an interesting subject.

Further, we studied the localization of the free scalar field and the free $U(1)$ gauge field. The spacetime geometry ensures that not only the scalar field but also the free $U(1)$ gauge field can be localized on the brane. Following the method in ref. \cite{Fu:2018erz}, we proved that there is no contradiction among the consistency conditions in our model. It means that a four-dimensional {free $U(1)$ gauge field} theory can be obtained by reducing the six-dimensional action. {
Localization of the free $U(1)$ gauge field implies that the four-dimensional Maxwell electromagnetic interaction can also be recovered on the brane. The KK modes of the $U(1)$ gauge field will make contributions to the Coulomb potential \cite{Guo:2011qt}, which deserves further study.}
In principle, the free $U(1)$ field can be localized on the brane for a $D$-dimensional model including compact extra dimensions.

Besides the localization of boson fields, the localization of fermion fields is still worth to explore. Whenever the dimension of the momentum space increases by two, the dimension of the spinor space will double. Different from the five-dimensional case, the structure of spinors will be more abundant in six-dimensional spacetime.


\acknowledgments

We thank Wen-Di Guo, Tao-Tao Sui, Ke-Yang, Bao-Min Gu, Yi Zhong, Hao Yu, Yu-Qiang Liu, Qin Tan, especially Yu-Peng Zhang and Chun-E Fu for many useful discussions. This work is supported in part by the National Key Research and Development Program of China (Grant No. 2020YFC2201503), the National Natural Science Foundation of China (Grants No. 11875151 and No. 12047501), and the 111 Project (Grant No. B20063).



\end{document}